%% file: ms.tex
\documentclass[sigconf,edbt]{acmart-edbt2024}

\def\BibTeX{{\rm B\kern-.05em{\sc i\kern-.025em b}\kern-.08em
    T\kern-.1667em\lower.7ex\hbox{E}\kern-.125emX}}
    
\usepackage{amsthm}
\usepackage{amsmath}
\usepackage{algorithmic}
\usepackage[linesnumbered,ruled,vlined]{algorithm2e} 
\usepackage{algorithm2e,setspace}
\usepackage{balance}
\usepackage{booktabs}
\usepackage{color}
\usepackage{colortbl}
\usepackage{caption}
\usepackage{capt-of}
\usepackage{enumitem}
\usepackage{enumitem,kantlipsum}
\usepackage{float}
\usepackage{graphicx}
\usepackage{hyperref}
\usepackage{multirow}
\usepackage{makecell}
\usepackage{subfigure}
\usepackage{textcomp}
\usepackage{xcolor}

\definecolor{gray}{rgb}{0.5,0.5,0.5}
\hypersetup{pdfauthor={Name}}

\newcommand{\rev}[1]{\textcolor[rgb]{0,0,0}{#1}}

\newcommand{\eatYou}[1]{}

\begin{document}

\title{DSPC: Efficiently Answering Shortest Path Counting on Dynamic Graphs (Full Version)}

\author{Qingshuai Feng}
\affiliation{%
  \institution{University of New South Wales}
  \streetaddress{High St Kensington}
  \city{Sydney}
  \state{NSW}
  \country{Australia}
  \postcode{2052}
}
\email{qsfeng@cse.unsw.edu.au}

\author{You Peng}
\affiliation{%
  \institution{University of New South Wales}
  \streetaddress{High St Kensington}
  \city{Sydney}
  \state{NSW}
  \country{Australia}
  \postcode{2052}
}
\email{unswpy@gmail.com}

\author{Wenjie Zhang}
\affiliation{%
  \institution{University of New South Wales}
  \streetaddress{High St Kensington}
  \city{Sydney}
  \state{NSW}
  \country{Australia}
  \postcode{2052}
}
\email{zhangw@cse.unsw.edu.au}

\author{Xuemin Lin}
\affiliation{%
  \institution{Shanghai Jiao Tong University}
  \city{Shanghai}
  \country{China}
}
\email{xuemin.lin@sjtu.edu.cn}

\author{Ying Zhang}
\affiliation{%
  \institution{University of Technology Sydney}
  \streetaddress{15 Broadway}
  \city{Sydney}
  \state{NSW}
  \country{Australia}
  \postcode{2007}
}
\email{ying.zhang@uts.edu.au}

\begin{abstract}
The widespread use of graph data in various applications and the highly dynamic nature of today's networks have made it imperative to analyze structural trends in dynamic graphs on a continual basis. The shortest path is a fundamental concept in graph analysis and recent research shows that counting the number of shortest paths between two vertices is crucial in applications like potential friend recommendation and betweenness analysis. However, current studies that use hub labeling techniques for real-time shortest path counting are limited by their reliance on a pre-computed index, which cannot tackle frequent updates over dynamic graphs. To address this, we propose a novel approach for maintaining the index in response to changes in the graph structure and develop incremental (\textsc{IncSPC}) and decremental (\textsc{DecSPC}) update algorithms for inserting and deleting vertices/edges, respectively. The main idea of these two algorithms is that we only locate the affected vertices to update the index. Our experiments demonstrate that our dynamic algorithms are up to four orders of magnitude faster processing for incremental updates and up to three orders of magnitude faster processing for hybrid updates than reconstruction.
\end{abstract}

\maketitle

\input{1_introduction}
\input{2_preliminaries}
\input{3_incremental}

\input{4_decremental}

\section{Experiments}

\subsection{Experiments Setup}

Experiments are carried out on a Linux server with Intel Xeon E3-1220 CPU and 520GB memory. All algorithms are implemented in C++ and complied with -O3 optimization. \rev{Each label entry $(v,d,c)$ is encoded in a 64-bit integer. Specifically, $v$, $d$, and $c$ take up 25, 10, and 29 bits, respectively. Labels of each vertex are stored in an array in descending order of ranking.}

\begin{table}[htb]
\centering
\caption{The Statistics of The Graphs}
  \begin{tabular}{|l|l|l|l|}
    \hline
    \cellcolor{gray!25}\textbf{Graph} & \cellcolor{gray!25}\textbf{Notation} & \cellcolor{gray!25}\textbf{n} & \cellcolor{gray!25}\textbf{m}\\
    \hline
     email-EuAll & EUA & 265,214 & 418,956 \\
    \hline
     NotreDame & NTD & 325,729 & 1,090,108 \\
    \hline     
     Stanford & STA & 281,903 & 1,992,636 \\
    \hline
     WikiConflict & WCO & 118,100 & 2,027,871 \\
    \hline
     Google & GOO & 875,713 & 4,322,051 \\
    \hline
     BerkStan & BKS & 685,231 & 6,649,470\\
    \hline
     Skitter & SKI & 1,696,415 & 11,095,298 \\
    \hline
     DBpedia & DBP & 3,966,924 & 12,610,982 \\
    \hline
     Wikilink War & WAR & 2,093,450 & 26,049,249 \\
    \hline
     \rev{Indochina-2004} & \rev{IND} & \rev{7,414,866} & \rev{150,984,819} \\
    \hline
\end{tabular}
\label{tab:graphs}
\end{table}

\subsubsection{Datasets}
\rev{Ten graphs from SNAP\footnote{https://snap.stanford.edu\label{snap}}, Konect\footnote{https://konect.cc\label{konect}} and LAW\footnote{https://law.di.unimi.it/\label{law}} are used for experiments and their details are shown in \autoref{tab:graphs}.} All graphs are undirected or converted to undirected. For incremental updates, 1,000 random edges are inserted into each graph.
To evaluate decremental updates, we randomly select $k$ edges from each graph for testing. $k$ is chosen in $\{50, 100\}$ according to the running time. For query performance evaluation, 10,000 random pairs of vertices are used for each graph and the average time is reported.

\begin{figure*}[htb]
	\begin{center}
        \subfigure[Incremental Update Time]{
			\label{inc_t}
			\centering
			\includegraphics[scale = 0.4]{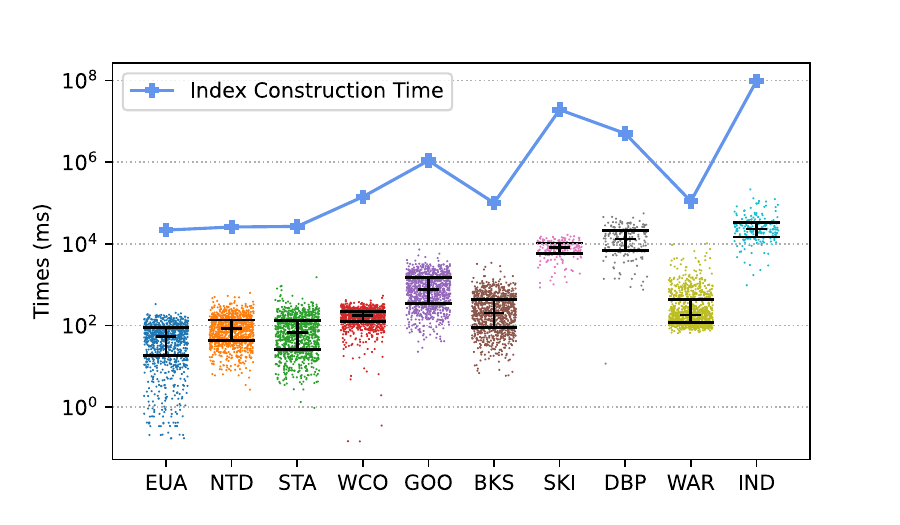} 
		} \hspace{-11mm}
		\subfigure[Decremental Update Time]{
			\label{dec_t}
			\centering
			\includegraphics[scale = 0.4]{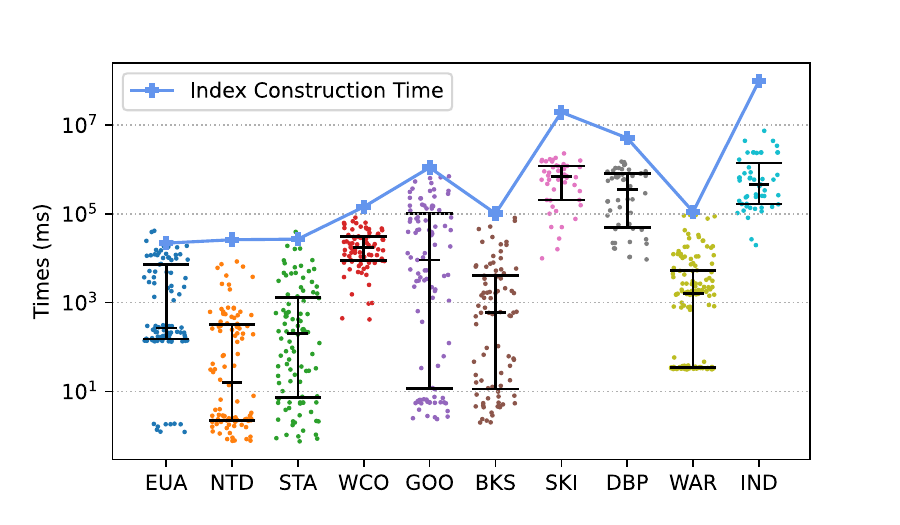} 
		} \hspace{-11mm}
		\subfigure[Query Time]{
			\label{query_t}
			\centering
			\includegraphics[scale = 0.4]{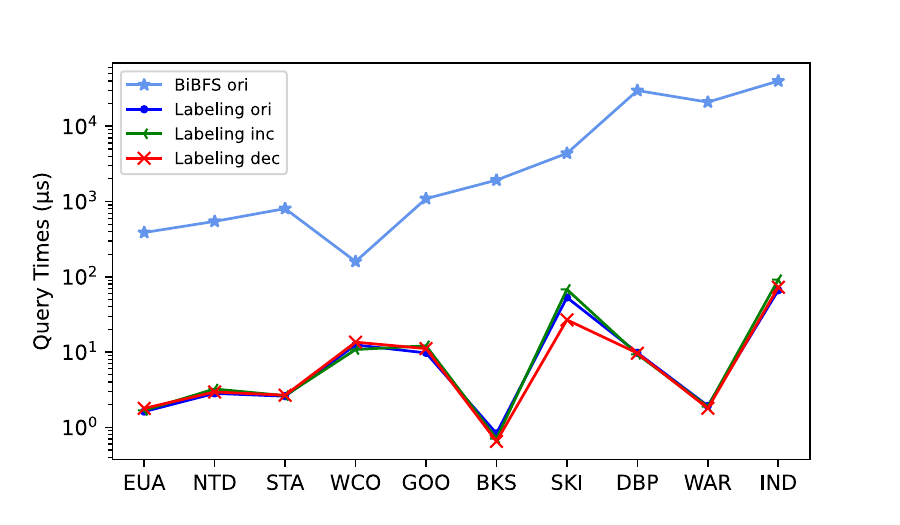}     
		}
	\end{center}
    \vspace{-3mm}
	\caption{Distribution of Running Times.} \label{fig:inc_dec_query}
\end{figure*}

\subsubsection{Algorithms}
 We use the building-from-scratch algorithm HP-SPC\cite{zhang2020hub} to compare with our incremental update algorithm \textsc{IncSPC} and the decremental algorithm \textsc{DecSPC}. For query evaluation, we compare the labeling query algorithm \textsc{SpcQUERY} against bidirectional BFS (BiBFS), which is an improved version of BFS for shortest path counting. The BiBFS algorithm conducts BFS searches from both query vertices and selects the side with the smaller queue size to continue each iteration until a common vertex from both sides is found. Lastly, accumulate the shortest path counting with minimum distance from all common vertices.

\begin{table}[htb]
\centering
\caption{Index Size (MB), Index Time and Average Inc/Dec Update Time (sec)}
  \begin{tabular}{|l|l|l|l|l|}
    \hline
    \cellcolor{gray!25}\textbf{Graph} & 
    \cellcolor{gray!25}\textbf{L Size} &
    \cellcolor{gray!25}\textbf{L Time} & 
    \cellcolor{gray!25}\textbf{\textsc{Inc}SPC} & \cellcolor{gray!25}\textbf{\textsc{Dec}SPC}\\
    \hline
     EUA & 353 & 21 & 0.05 & 4.59 \\
    \hline
     NTD & 578 & 26 & 0.1 & 0.56 \\
    \hline
     STA & 496 & 27 & 0.09 & 1.94 \\
    \hline
     WCO & 1,073 & 143 & 0.18 & 22.91 \\
    \hline
     GOO & 6,050 & 1,103 & 1.02 & 98.39 \\
    \hline
     BKS & 1,482 & 101 & 0.31 & 5.73 \\
    \hline
     SKI & 45,897 & 19,433 & 8.17 & 783.58 \\
    \hline
     DBP & 26,669 & 5080 & 15.49 & 480.37 \\
    \hline
     WAR & 1,595 & 111 & 0.43 & 8.41 \\
    \hline
     \rev{IND} & \rev{287,968} & \rev{98,011} & \rev{30.11} & \rev{1,058.34} \\
    \hline
\end{tabular}
\label{tab:update_time}
\end{table}

\subsection{Experiment Results of Incremental Update}

\subsubsection{Update Times}
Table \ref{tab:update_time}, columns 3 and 4, display the indexing times and average update times for HP-SPC and \textsc{IncSPC}, respectively. Furthermore, Figure \ref{fig:inc_dec_query}(a) presents a scatter plot showing the time distribution of each individual update. It includes the median, 25th percentile, and 75th percentile values, with the blue line indicating the index times. From the results, we observe that 
\rev{(i) On all tested graphs, the average running time of \textsc{IncSPC} is 2 to 4 orders of magnitude faster than HP-SPC (index reconstruction). For most graphs, the average update time falls within the range of 0.05 to 1.02 seconds, enabling real-time updates.
(ii) As the graph size increases, the disparity between update time and reconstruction time grows larger. Although \textsc{IncSPC} takes the longest time to update the largest graph, IND, which takes around 30 seconds for a single update. It is still much faster than reconstructing the entire index, which costs 98,011 seconds.
(iii) The update times exhibit a nearly symmetrical distribution within the interquartile range (between the 25th and 75th percentiles), with the median values serving as the center.
The above observations provide evidence that \textsc{IncSPC} is both stable and time-efficient in handling insertion scenarios. Especially for large graphs, using \textsc{IncSPC} can complete an update in seconds, avoiding the need to wait hours for a reconstruction.}

\subsubsection{Index Change}
\begin{figure}[t]
  \centering
  \includegraphics[scale=0.53]{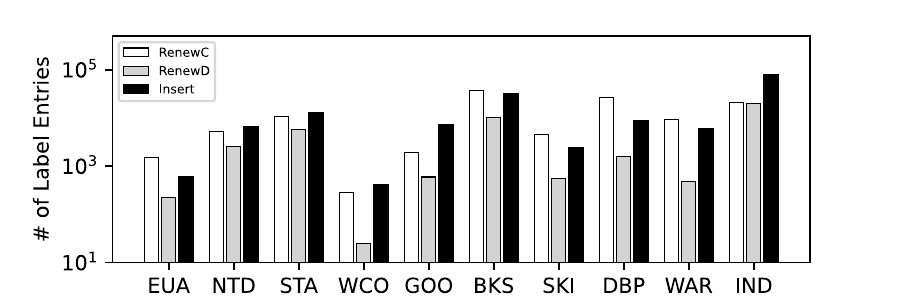}
  \caption{Average Number of Renewed Labels and Newly Inserted Labels for Incremental Update }
  \label{fig:inc_renew_insert}
\end{figure}

Figure \ref{fig:inc_renew_insert} compares the average number of updates for different types, such as \texttt{RenewC} (counting renewed only), \texttt{RenewD} (distance renewed), and \texttt{Insert} (newly inserted), respectively. We observe the following: 
(i) \texttt{RenewD} type of updates always make up the minority compared with the other two updates in all graphs. A possible explanation may be the inherent property of the graphs, i.e., a new edge may generate more shortest paths with unchanged distances. 
(ii) The magnitude of \texttt{Insert} also illustrates the average index increase for incremental updates. \rev{Considering that each label entry takes up 8 bytes, the average increase in index size ranges from 3KB to 105KB for most graphs and it is 646KB for the largest graph IND.} Nevertheless, the change is minor in contrast to the original index size (Column 2 in Table~\ref{tab:update_time}). 
These findings confirm that despite \textsc{IncSPC} keeping some redundant labels, it is still space efficient in terms of index change size.

\subsection{Experiment Results of Decremental Update}

\subsubsection{Update Times}
Table \ref{tab:update_time} column 5 shows the average update time taken by \textsc{DecSPC}. Figure \ref{fig:inc_dec_query}(b) illustrates the detailed update time of each \textsc{DecSPC} execution. For graph SKI, DBP, and IND, a small number of tests run too long and have been set as a timeout. From the results, we observe the following: 
(i) The time cost of \textsc{DecSPC} is longer than \textsc{IncSPC} but still 1 to 2 orders of magnitude faster than reconstruction. In a few cases, the update time is comparable to the time of reconstruction.
(ii) We observe major differences in the distribution of the running time between \textsc{IncSPC} and \textsc{DecSPC}, the latter is more dispersed. This discrepancy is attributed to the inherent complexity of \textsc{DecSPC}. 
(iii) It is apparent from Figure \ref{fig:inc_dec_query}(b) that some updates complete much faster than others, e.g., the bottom dots in graph ENU and WAR. This is because the optimization of isolated vertex deletion reduces the running time significantly.
In certain real-world dynamic networks like e-mail networks and co-author networks, vertex/edge deletion is uncommon. Additionally, insertion updates significantly outnumber deletion updates in many other networks~\cite{akiba2014dynamic,d2019fully}. Consequently, insertion updates dominate hybrid updates, resulting in a lower average update time compared to deletion updates. In summary, both \textsc{IncSPC} and \textsc{DecSPC} offer practical solutions to handle various update types, effectively avoiding the need for reconstruction.

\begin{figure}[t]
  \centering
  \includegraphics[scale=0.53]{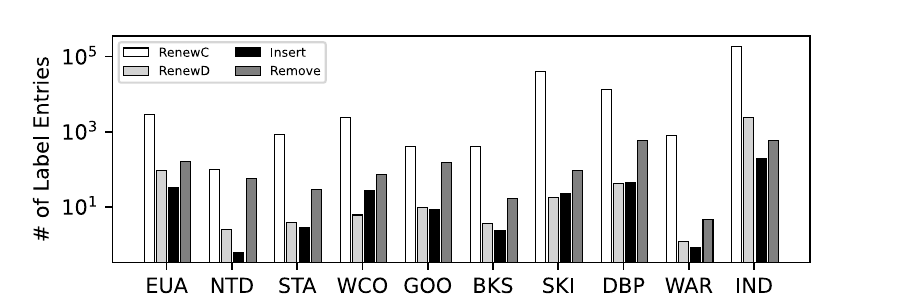}
  \caption{Average Number of Renewed Labels, Newly Inserted Labels, and Removed Labels for Decremental Update}
  \label{fig:dec_cdir}
\end{figure}

\subsubsection{Index Change}
Figure \ref{fig:dec_cdir} shows the average number of each type of label updates taken by \textsc{DecSPC}. Unlike \textsc{IncSPC}, \textsc{DecSPC} includes additional removed labels (\texttt{Remove}). The results indicate that:
(i) Renewed labels account for the majority of all kinds of updates, especially for \texttt{RenewC}, but they don't induce a change in the index size.
(ii) The average reduction in index size is the difference between \texttt{Remove} and \texttt{Insert}, and this magnitude is only in the range of a few kilobytes.
These results indicate that \textsc{DecSPC} is space efficient, i.e., a minor change in index size occurs after the decremental updates.

\begin{figure}[t]
	\begin{center}
        \hspace{-6mm}
        \subfigure[BKS]{
			\label{bks_stream}
			\centering
			\includegraphics[scale = 0.215]{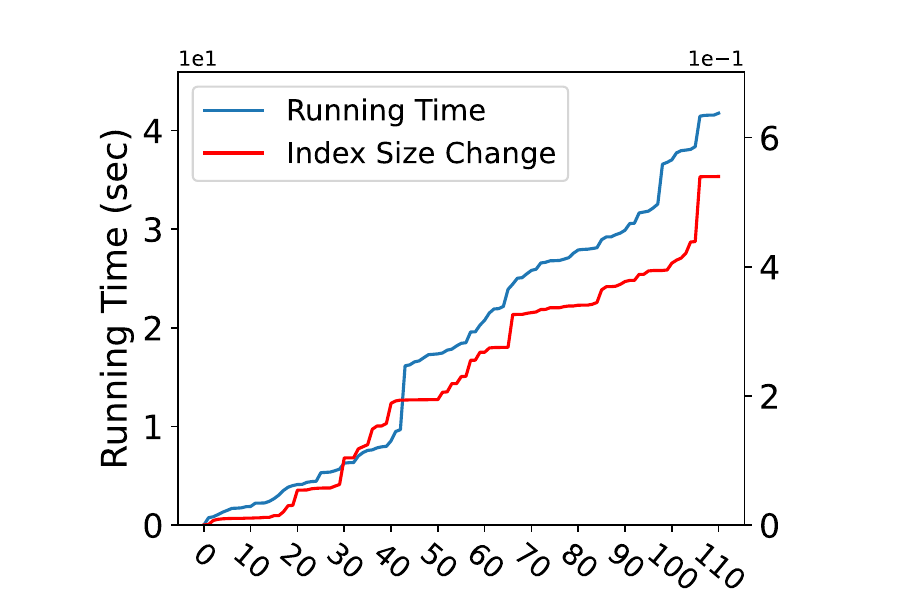} 
		} \hspace{-11.2mm}
		\subfigure[WAR]{
			\label{war_stream}
			\centering
			\includegraphics[scale = 0.215]{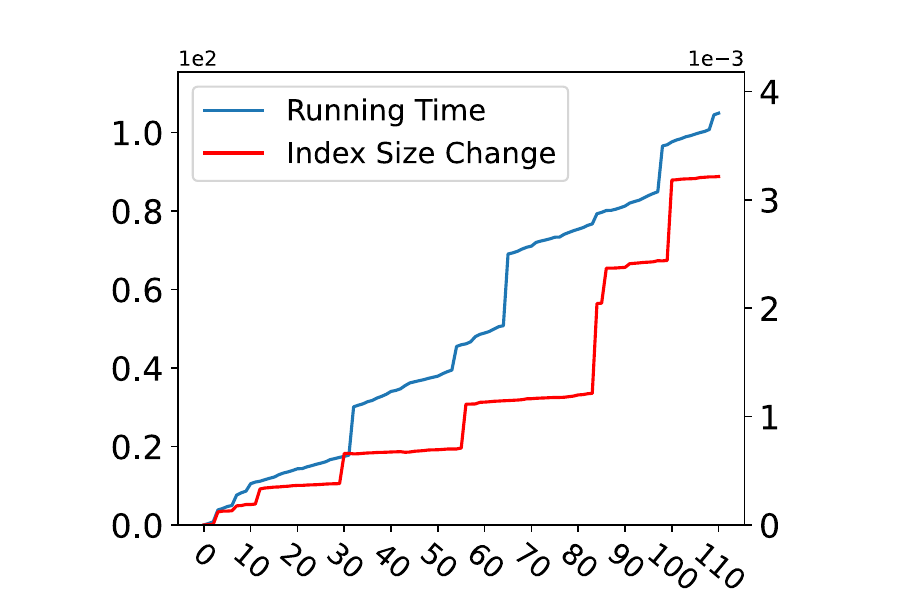} 
		} \hspace{-11.2mm}
		\subfigure[IND]{
			\label{ind_stream}
			\centering
			\includegraphics[scale = 0.215]{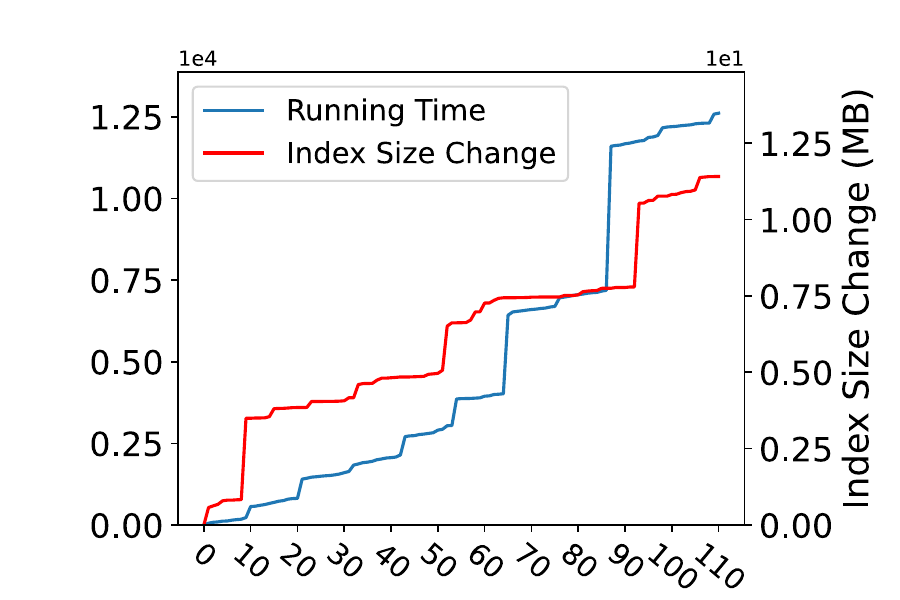}
		}
	\end{center}
    \vspace{-3mm}
	\caption{\rev{Accumulated Running Times (sec) and Index Size Changes (MB) of Streaming Update.}} \label{fig:streaming_update}
\end{figure}

\subsection{\rev{Experiment Results of Streaming Update}}
\rev{We also conduct experiments with streams of updates. In particular, we considered a stream of 100 random edge insertions and 10 random edge deletions on graphs BKS, WAR, and IND. \autoref{fig:streaming_update} illustrates the accumulated running times (blue lines) of the streaming updates as well as the index changes (red lines) for each graph. In most cases, the running times increase gradually, except for a few cases where edge deletion updates are particularly time-consuming, as shown by sudden increases in the figures. The average running times for hybrid streaming updates on graphs BKS, WAR, and IND are 0.37, 0.95, and 114.64 seconds, respectively. These times are still 2 to 3 orders of magnitude faster than reconstructing everything. Furthermore, the total index size increase due to the streaming updates is minimal, with only 539KB, 3KB, and 11.4MB for the three tested graphs, respectively. This increase is negligible when compared to the original size of the index.}

\subsection{\rev{Experiment Results of Skewed Update}}

\begin{figure}[t]
	\begin{center}
        \subfigure[BKS (\textsc{IncSPC})]{
			\label{bks_skewed_inc}
			\centering
			\includegraphics[scale = 0.205]{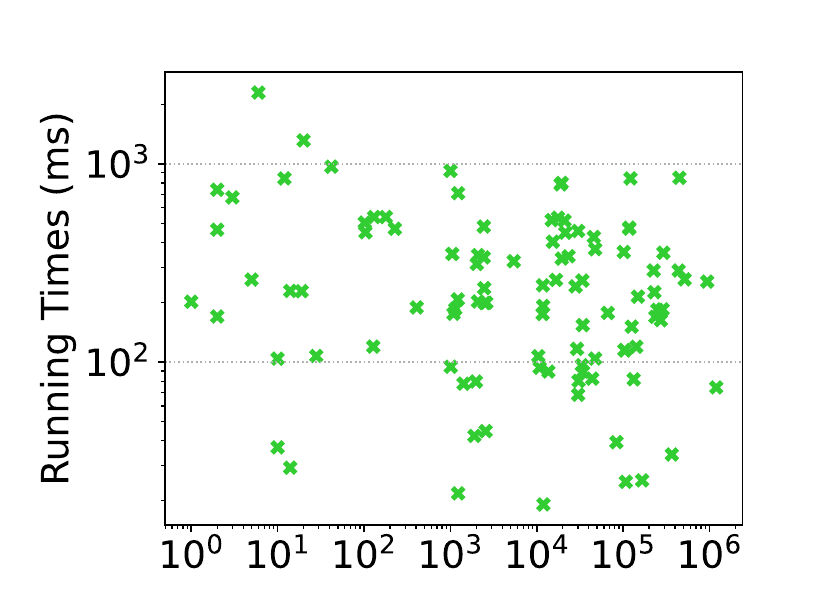} 
		} \hspace{-7.8mm}
		\subfigure[WAR (\textsc{IncSPC})]{
			\label{war_skewed_inc}
			\centering
			\includegraphics[scale = 0.205]{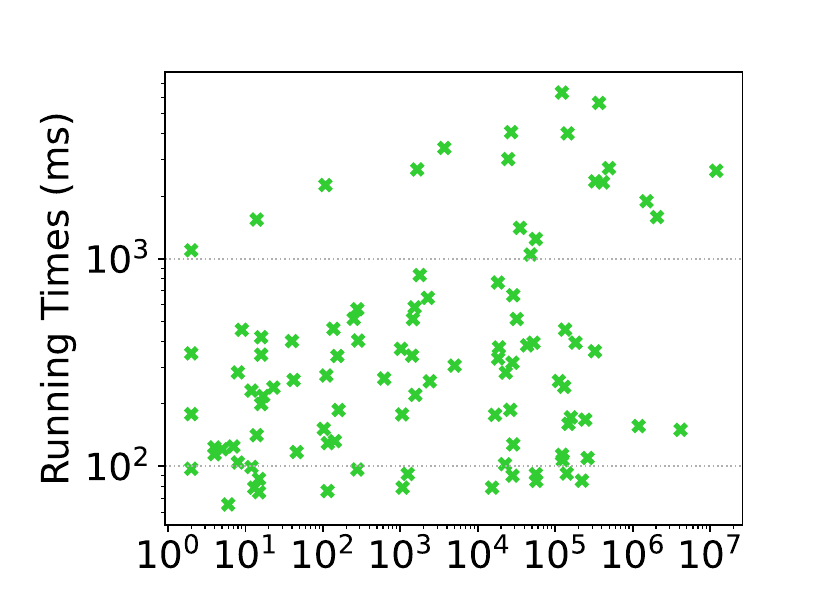} 
		} \hspace{-7.8mm}
		\subfigure[IND (\textsc{IncSPC})]{
			\label{ind_skewed_inc}
			\centering
			\includegraphics[scale = 0.205]{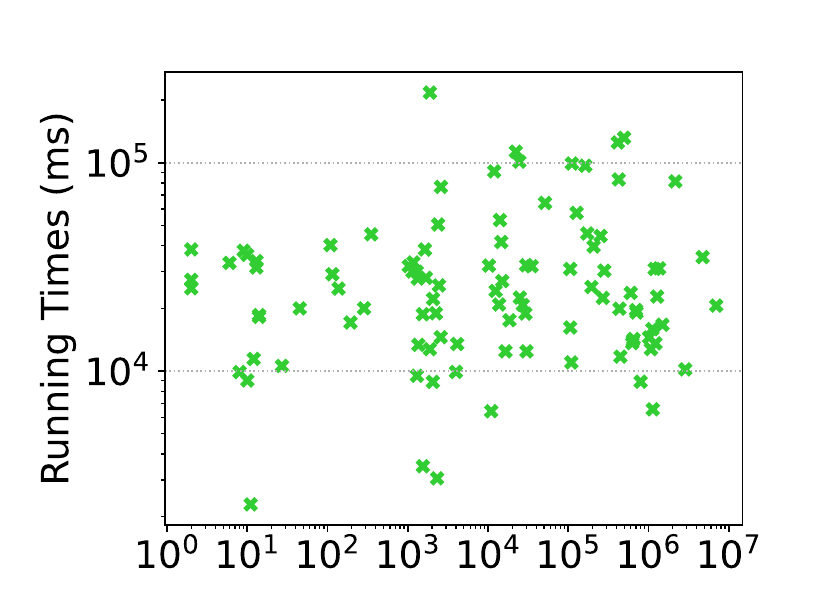}
		}\vspace{-2mm}
        \subfigure[BKS (\textsc{DecSPC})]{
			\label{bks_skewed_dec}
			\centering
			\includegraphics[scale = 0.205]{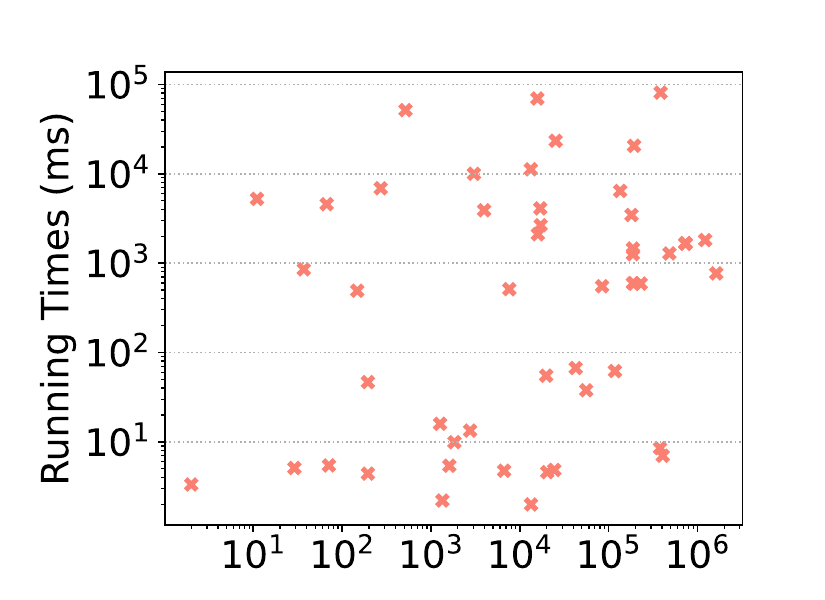} 
		} \hspace{-7.8mm}
		\subfigure[WAR (\textsc{DecSPC})]{
			\label{war_skewed_dec}
			\centering
			\includegraphics[scale = 0.205]{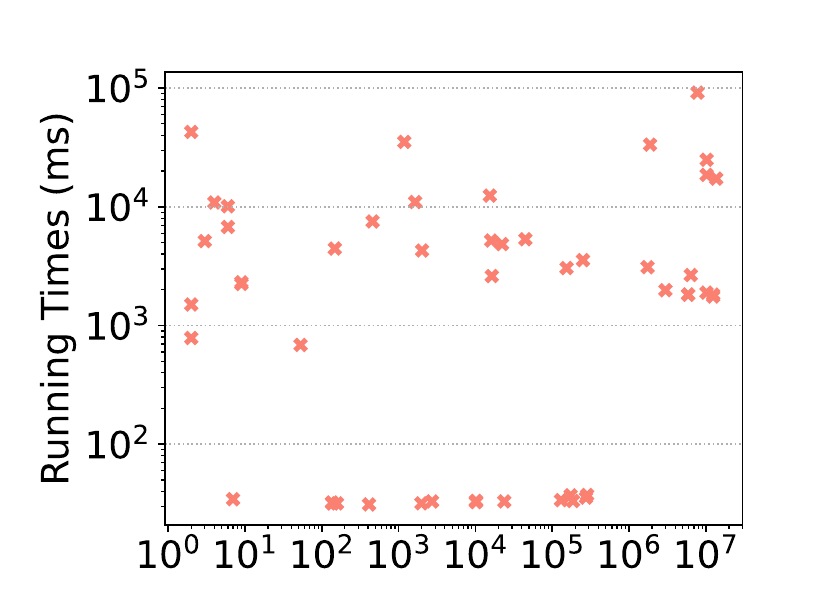} 
		} \hspace{-7.8mm}
		\subfigure[IND (\textsc{DecSPC})]{
			\label{ind_skewed_dec}
			\centering
			\includegraphics[scale = 0.205]{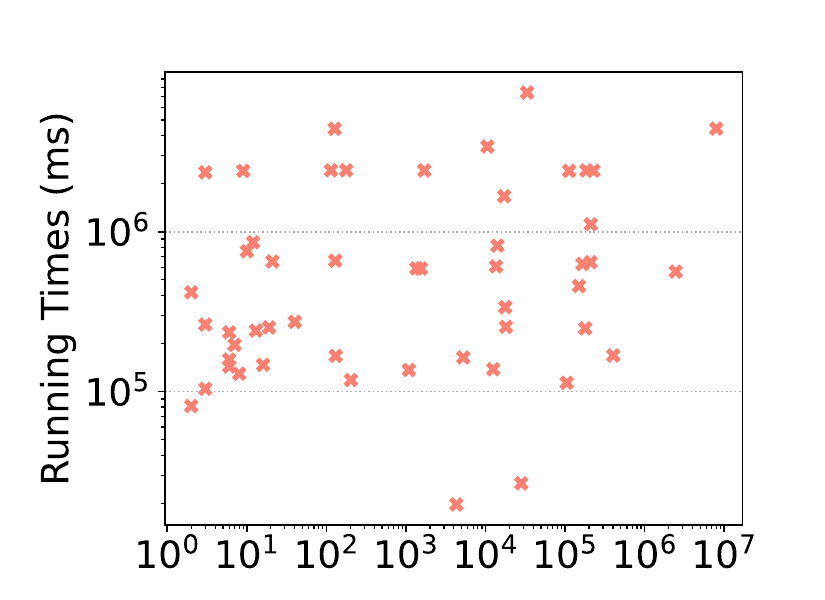}
		}
	\end{center}
    \vspace{-3mm}
	\caption{\rev{Running Times (ms) of \textsc{IncSPC} and \textsc{DecSPC} (Varying Degrees of Inserted Edges and Deleted Edges).}} \label{fig:skewed_inc}
\end{figure}

\rev{In this experiment, we aim to investigate the distribution of running times of the algorithms for varying degrees of inserted and deleted edges. The degree of an edge $(u,v)$ is defined as \texttt{deg}($u$)*\texttt{deg}($v$). Experiments are performed on graphs BKS, WAR, and IND by adding 100 edges and deleting 50 edges with varying degrees. The running times of \textsc{IncSPC} and \textsc{DecSPC} are depicted in \autoref{fig:skewed_inc}. The results indicate that there is no significant correlation between the degree of the tested edges and the running times of the algorithms, which is consistent with their time complexities. \textsc{IncSPC}'s running time is proportional to the number of visited vertices during the BFS, while \textsc{DecSPC}'s running time increases as the size of the affected vertex sets. In fact, there is no strong correlation between the degree of the tested edges and these two metrics. Despite having a low degree, an edge may still have a high number of shortest paths passing through it.
}

\subsection{Experiment Results on sets SR and R}
Table \ref{tab:size_srr} compares the cardinalities of $SR$ and $R$ in decremental experiments. Because the edges are undirected, we have $(a,b)\equiv (b,a)$ for decremental updates. We let $SR_a$ accommodate the set with more affected hubs. $SR_a$ and $SR_b$ are swapped if $|SR_b|>|SR_a|$. The results demonstrate that the number of affected hubs, $|SR|=|SR_a|+|SR_b|$, is much smaller compared to $|R|$ which equals $|R_a|+|R_b|$. It proves the effectiveness of \textsc{DecSPC} in selecting a limited number of affected hubs and thereby minimizing unnecessary computations.

\begin{table}[t]
\centering
\caption{Average size of $SR_a, SR_b, R_a, R_b$}
  \begin{tabular}{|l|l|l|l|l|}
    \hline
    \cellcolor{gray!25}\textbf{Graph} & \cellcolor{gray!25}\textbf{$SR_a$} & \cellcolor{gray!25}\textbf{$SR_b$} & \cellcolor{gray!25}\textbf{$R_a$} 
    & \cellcolor{gray!25}\textbf{$R_b$} \\
    \hline
     EUA & 115,328 & 19 & 52,701 & 2,490 \\
    \hline
     NTD & 68,590 & 2 & 23,740 & 3,488 \\
    \hline
     STA & 435 & 7 & 48,076 & 15,332 \\
    \hline
     WCO & 3,153 & 28 & 30,269 & 6,709 \\
    \hline
     GOO & 53,261 & 10 & 129,767 & 1,471 \\
    \hline
     BKS & 1,434 & 5 & 183,326 & 10,018 \\
    \hline
     SKI & 4,810 & 25 & 492,358 & 93,017 \\
    \hline
     DBP & 480,053 & 109 & 1,094,036 & 388,126\\
    \hline
     WAR & 73,534 & 2 & 500,393 & 219,165 \\
    \hline
     \rev{IND} & \rev{35,397} & \rev{15} & \rev{1,315,357} & \rev{158,157} \\
    \hline
\end{tabular}
\label{tab:size_srr}
\end{table}

\subsection{Experiment Results on Query Time}
Figure \ref{fig:inc_dec_query}(c) illustrates the query time for BiBFS and labeling methods, utilizing indexes for the static graph (ori) as well as after incremental (inc) and decremental (dec) updates. As expected, labeling methods outperform the online algorithm BiBFS in query evaluation, exhibiting query times up to four orders of magnitude faster. Comparing the results of labeling methods, both \textsc{IncSPC} and \textsc{DecSPC} exhibit consistent query times, indicating minimal impact from the update algorithms on the index size.

\section{Related Work}

\noindent\textbf{Pattern Counting.} Counting specific patterns in graphs is a fundamental problem in graph theory, often falling under the \#P-complete complexity class. Examples include counting simple paths between two vertices \cite{jerrum1994counting} and counting perfect matchings in bipartite graphs \cite{fukuda1994finding}. Counting paths or cycles with length constraint, parameterized by k, is \#W[1]-complete. Other studies explore counting shortest paths on planar graphs \cite{bezakova2018counting}, probabilistic biological networks \cite{ren2018shortest}, and providing approximate counts of shortest paths in directed acyclic graphs \cite{mihalak2016approximately}. In addition to counting paths or cycles, \cite{jain2017fast} and \cite{pinar2017escape} introduce randomized algorithms that offer proven guarantees for counting cliques and 5-vertex subgraphs within a given graph, respectively.

\vspace{1mm}

\noindent\textbf{2-hop Labeling on Static Graph.} 2-hop labeling is an index-based method that is widely used to evaluate shortest distances on static graphs. Each vertex is assigned a label set. The shortest distance between any pair of vertices can be calculated through their label sets instead of traversing the graph. For road networks with small tree-widths, \cite{ouyang2018hierarchy, chen2021p2h} propose hierarchical 2-hop indexes based on the hierarchy of vertices. For real scale-free graphs, Pruned Landmark Labeling (PLL) \cite{akiba2013fast} is the state-of-the-art schema to construct the distance index and has many variants. For example, \cite{li2019scaling} introduces a parallel version of PLL and \cite{li2020scaling} combines the hierarchical index and PLL index on different parts of the graph for scaling up. A recent study \cite{zhang2020hub} extended PLL to address the shortest path counting problem by supplementing additional labels and counting information, which provides the underlying SPC-Index used in this work. 
\rev{Another 2-hop labeling for shortest distance queries is Highway Cover Labeling~\cite{farhan2018highly}. Unlike PLL, Highway Cover Labeling employs a partial index, utilizing only a small number of vertices as hubs. Its query evaluation involves computing a distance upper bound using the index followed by performing a distance-bounded BFS. }

\vspace{1mm}

\noindent\textbf{Index Maintenance for 2-hop Labeling.} The most common changes to the graph are vertex/edge insertion and deletion. \rev{Many works attempt to maintain various 2-hop shortest-distance indexes for the above cases and there is a lack of research on the maintenance of SPC-Index.} For the SD-Index under PLL, outdated labels will not affect any shortest distance when a new edge is inserted. Based on this lemma, \cite{akiba2014dynamic} proposes an incremental update algorithm for the edge insertion case, but the updated index loses the minimal property by considering the time cost trade-off. In case of an edge is deleted, outdated labels will underestimate the shortest distances if they are not removed. To this end, \cite{d2019fully, qin2017efficient} use a similar partial reconstruction strategy to handle vertex/edge deletion, but the update time is not practical on some graphs. Propagation-based update algorithms for weighted graphs are devised in \cite{zhang2021efficient, chen2021p2h}.

\rev{Recently, \cite{farhan2022fast} proposed a fully dynamic algorithm for updating the Highway Cover Labeling, which was later extended by \cite{farhan2022batchhl} to support batch updates. As a result of the disparity in the labeling schema, the update strategies of these methods cannot be applied to the maintenance of the SPC-Index. \cite{zhang2021dynamic} presented an update algorithm for maintaining the hierarchy-based shortest distance index proposed by \cite{ouyang2018hierarchy}, without the need to traverse the graph. However, their method is specific to their own labeling schema and cannot be applied to other labeling schemas. \cite{qiu2022efficient} proposed a shortest path counting index based on \cite{ouyang2018hierarchy}'s hierarchy-based labeling, but its applicability is also limited to road networks as with \cite{ouyang2018hierarchy}. This is due to the fact that scale-free networks, such as web graphs, often have a dense core, making the tree decomposition to the core infeasible \cite{ouyang2018hierarchy}. Additionally, there is currently no update algorithm available in the literature for \cite{qiu2022efficient}'s shortest path counting index in road networks.}

\section{Limitations of This Work}
\noindent\textbf{\rev{Parallel Implementation.}} \rev{Parallel algorithms play a pivotal role in accelerating computations and improving scalability for graph processing. However, developing a parallel update algorithm for SPC-Index poses challenges due to the strict order dependency among all vertices. Specifically, when dealing with an affected hub $h$, it is critical to ensure the correctness of all the labels $(h',d,c)$ with $h'\leq h$, which forces the affected vertices to be updated sequentially. Nevertheless, there is a possibility of parallel implementation within the update process for each affected hub. During the BFS rooted at an affected hub, vertices at the same distance level, denoted by $D[\cdot]$ in Algorithms~\ref{alg:inc_upd} and \ref{alg:dec_update}, can be tested and updated simultaneously because they all rely on the identical and partial up-to-date index. The algorithm proceeds to the next distance level only when updates for all vertices at the current distance level are completed. The above idea is theoretically feasible in a shared-memory system as each process requires access to the index and the ability to share information to compute the path count for vertices at the next level. However, due to various factors such as the skewed nature of the graph, it is not guaranteed that there will always be significant improvement in performance. Hence, the design of an effective parallel update algorithm for SPC-Index is challenging in both shared-memory and distributed-memory systems.}

\vspace{1mm}

\noindent\textbf{\rev{Vertex Ordering Changes.}} \rev{The vertex ordering is crucial for pruned landmark labeling, as it has a significant impact on indexing time, index size, and query time. However, the initial vertex ordering may become irrelevant after a series of updates. For example, under degree-based ordering, a vertex with a low degree at first will still have a low rank even if its degree increases later. Changing the ordering to an index is challenging since it may result in a large number of labels that violate the rank constraint. One possible solution is to use the lazy strategy, i.e., reconstructing the entire index after a certain number of updates. It remains an open problem of how to efficiently renew the index when the ordering changes.}

\section{Conclusion}
\rev{In this paper, we study the maintenance of the 2-hop labeling for shortest path counting. We adopt the SPC-Index built by HP-SPC as the underlying index and aim to maintain the index regarding the topological changes to the graph. We propose two novel dynamic algorithms for SPC-Index, including incremental algorithm \textsc{IncSPC} and decremental algorithm \textsc{DecSPC}, and also optimization for the isolated vertex. To the best of our knowledge, our proposed method is the first approach for updating SPC-Index. We have empirically verified the effectiveness and efficiency of our approach, which can process incremental updates up to four orders of magnitude faster and hybrid updates up to three orders of magnitude faster compared to reconstructing the index. In the future, we intend to investigate parallel and distributed algorithms to further improve the efficiency and scalability of our approach.}

\newpage

\bibliographystyle{ACM-Reference-Format}
\bibliography{mybib}

\newpage

\appendix
\section{Application Example}

\begin{figure}[htb]
    \centering
    \includegraphics[scale = 0.55]{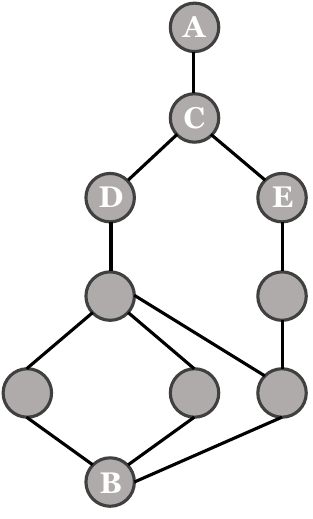}
    \caption{Shortest paths in the collaboration network between the two scientists labeled A and B}
    \label{fig:app_collab}
\end{figure}

In scientific collaboration networks, two scientists are deemed connected when they have collaborated as coauthors on a research paper. Figure~\ref{fig:app_collab} shows the shortest paths of collaborations in the Los Alamos Archive and illustrates the connections between two colleagues of C, denoted by the labeled vertices A and B. Although both scientists are involved in researching social networks of different types, the shortest path between them does not solely rely on collaborations within their specific fields. Notably, the connections between C and D, as well as E, are formed through papers on topics unrelated to networks or graph theory. Thus, a certain number of shortest paths between two scientists may suggest potential future collaborations, even though the scientists connected in the middle are not in the relevant fields. As global scientific collaborations become increasingly prevalent, having dynamic access to information such as pairwise shortest path counting becomes highly valuable.

\section{Proofs}

\subsection{Proof of \autoref{theorem:inc_correct}}

\begin{figure} [htb]
  \centering
  \includegraphics[scale=0.13]{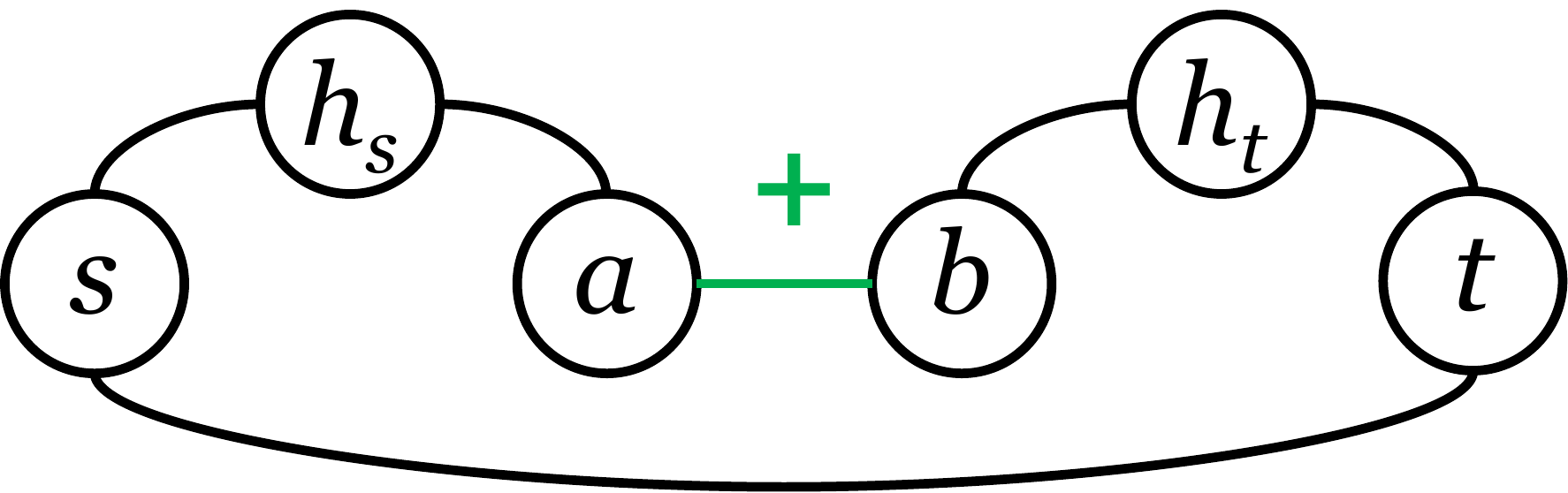}
  \vspace{-2mm}
  \caption{Example of $(a,b)$ Insertion.}
  \label{fig:inc_proof}
  \vspace{-4mm}
\end{figure}

\begin{proof}
    To verify the cover constraint, we need to prove that the shortest path counting between any vertex pair $(s, t)$ can be correctly computed. As shown in Figure~\ref{fig:inc_proof}, edge $(a,b)$ is inserted. Assume $s$ and $t$ are two arbitrary vertices, and there exist new shortest paths between them over $(a,b)$. We use $\texttt{sd}_{i+1}(s,t)$ to denote their new distance. The bottom line between $s$ and $t$ illustrates the shortest paths between them before the edge insertion, and we use $\texttt{sd}_i(s,t)$ to represent the distance. \textsc{IncSPC} will keep the existing counting if $\texttt{sd}_i(s,t)=\texttt{sd}_{i+1}(s,t)$. The reason is when new shortest paths of the same length as before are found, we accumulate the counting to the existing label or insert a new label. Following that, we prove all new shortest paths between $s$ and $t$ induced by $(a,b)$ counted and added to the index. Assume path $s\rightarrow h_s\rightarrow a\rightarrow b\rightarrow h_t\rightarrow t$ represents a new shortest path between $(s,t)$ over $(a,b)$. We use $h_s$ to denote the hub of the sub-path from $s$ to $a$ and $h_t$ be the hub of the sub-path from $b$ to $t$. Thus, $h_s$ has the highest rank along the sub-path from $s$ to $a$ and $h_t$ has the highest rank along the sub-path from $b$ to $t$. Assume $h_s\leq h_t$ without loss of generality. Then, $h_s$ has the highest rank along the new shortest path between $(s,t)$ and $h_s\in L_i(a)$. In \textsc{IncSPC}, $h_s$ will perform BFS from $b$ and count all new shortest paths to visited vertices until meeting an unaffected vertex. In addition, all possible affected hubs are contained in $L_i(a)\cup L_i(b)$ s.t. \textsc{IncSPC} will count all new shortest paths to the index. Hence, $L_{i+1}$ updated by \textsc{IncSPC} will obey ESPC for the graph $G_{i+1}$.
\end{proof}

\subsection{Proof of \autoref{theorem:inc_time}}

\begin{proof}
    There are $2\cdot l$ pruned BFS at most rooted at $\texttt{AFF}=\{h|h\in L_i(a)\cup L_i(b)\}$. During the pruned BFS, for each visited vertex, a SPC query is evaluated which costs $O(l)$ time. Thus, the overall time is $O(2\cdot l \cdot k \cdot l)$ which is $O(kl^2)$.
\end{proof}

\subsection{Proof of Lemma~\ref{lemma:label_update}}

\begin{proof}
    We first prove that i) only the label entries $(v,\cdot,\cdot)$ may require updates or to be added if $v\in SR$ and ii) the labels in $L(u)$ may require updates if $u\in SR\cup R$. Then we prove the above lemma. The update mentioned in the above lemma includes the change of distance or counting element, or the label entry is supposed to be deleted. The reason is that for any vertex $v$, it will not be affected if $\texttt{sd}_i(v,a)=\texttt{sd}_i(v,b)$. We exclude this case in the following proof and assume $|\texttt{sd}_i(v,a)-\texttt{sd}_i(v,b)|=1$ without loss of generality. 

    We prove statement i) by contradiction. According to the definition of $SR$ set, a vertex $v$ belongs to $SR$ if $v\in L_i(a)\cap L_i(b) \vee \texttt{spc}_i(v,a)=\texttt{spc}_i(v,b)$. Suppose there exists a vertex $v\in SR$, i.e., there exists $(v,\cdot,\cdot)$ requires update or to be added, and $v\notin L_i(a)\cap L_i(b) \wedge \texttt{spc}_i(v,a)\not=\texttt{spc}_i(v,b)$. Assume $\texttt{sd}_i(v,a)+1=\texttt{sd}_i(v,b)$, we have $\texttt{spc}_i(v,b)>\texttt{spc}_i(v,a)$ because all $\texttt{sp}_i(v,a)$ concatenate $(a,b)$ will be $\texttt{sp}_i(v,b)$ and there exists other $\texttt{sp}_i(v,b)$ not passing through $(a,b)$. We use $\texttt{sp}^{a\cdot b}_i(v,b)$ and $\texttt{sp}^{a/b}_i(v,b)$ to denote a path in former case and latter case, respectively. For $v\notin L_i(a)\cap L_i(b)$, we first check $v\notin L_i(a)\wedge v\in L_i(b)$. In this case, all $\texttt{sp}^{a\cdot b}_i(v,b)$ are covered by hubs other than $v$. Hence, $(v,\cdot,\cdot)\in L_i(b)$ is not affected and neither other vertices after $b$. Then for $v\in L_i(a)\wedge v\notin L_i(b)$, we have $\texttt{spc}_i(\Hat{v},b)=0$ and for all $(v,\cdot,\cdot)$ in $L_i$, none of them cover a path going through $(a,b)$. Thus, no $(v,\cdot,\cdot)$ requires update. In addition, because $\texttt{spc}_i(v,a)\not=\texttt{spc}_i(v,b)$, the distance from $v$ to all vertices will not change. No new labels with $v$ as hub will be added. The proof for $\texttt{sd}_i(v,b)+1=\texttt{sd}_i(v,a)$ is symmetric. Therefore, no $(v,\cdot,\cdot)$ requires update or to be added. Contradiction.

    For statement ii), $SR\cup R$ contains all vertices of which $|\texttt{sd}_i(v,a)-\texttt{sd}_i(v,b)|=1$. All other vertices will not be affected by $(a,b)$, i.e., they have no shortest path passing through $(a,b)$. Thus, statement ii) holds.

    Finally, because for any two vertices $v$ and $u$ such that $\texttt{sd}_i(v,a)+1=\texttt{sd}_i(v,b)$ and $\texttt{sd}_i(u,a)+1=\texttt{sd}_i(u,b)$, $\texttt{sd}_i(v,u)=\texttt{sd}_{i+1}(v,u)$ holds, the shortest paths between $v$ and $u$ will not be affected by the deletion of $(a,b)$. The conclusion can also be achieved if $\texttt{sd}_i(v,b)+1=\texttt{sd}_i(v,a)$ and $\texttt{sd}_i(u,b)+1=\texttt{sd}_i(u,a)$. Now, we can prove that Lemma~\ref{lemma:label_update} holds.
\end{proof}

\subsection{Proof of \autoref{theorem:dec_correct}}

\begin{proof}
    We prove the correctness by contradiction. Assume the label $(v,d,c)\in L_{i+1}(u)$ is incorrect which causes one or more queries incorrect. Thus, we have $v\in SR_a$ and $u\in SR_b\cup R_b$ (Or $v\in SR_b$ and $u\in SR_a\cup R_a$) by Lemma~\ref{lemma:label_update}. As the affected hubs are processed in descending order of rank. When processing hub $v$, we ensure labels $(\Bar{v},\cdot,\cdot)$ are correct for all $\Bar{v}$ such that $\Bar{v}\leq v$. This implies the distance $d$ returned by \textsc{PreQUERY}$(v,u)$ has $d\geq \texttt{sd}_{i+1}(v,u)$. The BFS starts from $v$ and finds the shortest distance $D[u]$ and shortest path counting $C[u]$ when meeting $u$ because of the property of BFS. With the ranking constraint, $C[u]=\texttt{spc}_{i+1}(\Hat{v},u)$ when $D[u]=\texttt{sd}_{i+1}(v,u)\leq d$. In this case, the test of line 8 in Algorithm~\ref{alg:dec_update} succeeds and the algorithm reaches line 9. Now, $(v,D[u],C[u])$ is inserted into $L(u)$ or updated. The BFS stops when meeting a vertex $w$ such that $\texttt{spc}_{i+1}(\Hat{v},w)=0$ so that it visits all vertices $u$ with $\texttt{spc}_{i+1}(\Hat{v},u)>0$. After the BFS, the removal phase will delete all $(v,\cdot,\cdot)$ such that the label is dominated by higher-ranked hubs or the owner of the label is disconnected with $v$. Therefore, $(v,d,c)\in L_{i+1}(u)$ is correct after Algorithm~\ref{alg:dec_spc}.
\end{proof}

\subsection{Proof of \autoref{theorem:dec_time}}

\begin{proof}
    Algorithm~\ref{alg:srr_search} visits $O(S_a+S_b)$ vertices in total to find $SR$ and $R$, and each time visiting a vertex, an SPC query is evaluated which costs $O(l)$ time. Thus, its time complexity is $O((S_a+S_b)l)$.
    Algorithm~\ref{alg:dec_update} conducts $O(s_a)$ (or $O(s_b)$) BFSs rooted at hubs from $SR_a$ (or $SR_b$) and visits $O(k)$ vertices for each BFS. Each time visiting a vertex, a \textsc{PreQUERY} is evaluated which costs $O(l)$ time. Then $O(S_b)$ (or $O(S_a)$) vertices are checked for label removal and it costs $O(l)$ for each vertex. Thus, its time complexity is $O(s_a(kl+S_bl)+s_b(kl+S_al)$. And the total time complexity for \textsc{DecSPC} is $O((S_a+S_b)l)+O(s_a(kl+S_bl)+s_b(kl+S_al)=O((S_a+S_b)l+(k+S_b)ls_a+(k+S_a)ls_b)$.
\end{proof}

\nobalance
\section{Extensions}

\subsection{Extension To Directed Graphs}
To extend the underlying HP-SPC index for a directed graph, The major modification is that instead of assigning one label set $L(v)$ for $v$ as in undirected graph, two label sets for each vertex, $L_{in}(v)$ and $L_{out}(v)$, are required. The labels in $L_{in}(v)$ represent the shortest paths from other vertices to $v$, while the labels in $L_{out}(v)$ represent the shortest paths from $v$ to other vertices. Each label $(h, d, c)$ in $L_{in}(v)$ signifies the existence of $c$ shortest paths with distance $d$ from hub $h$ to $v$, where $h$ is the highest-ranked vertex. Similarly, the labels in $L_{out}(v)$ represent the shortest paths from $v$ to other vertices. The index construction involves performing two BFSs from each hub, one in each direction, to generate labels for the $L_{in}$ and $L_{out}$ sets of other vertices. To answer $SPC(s,t)$ query, simply scan through $L_{out}(s)$ and $L_{in}(t)$ like \textsc{SpcQUERY}.

For incremental updates to the directed SPC-Index, assume edge $(a,b)$ is inserted. In this case, the affected hubs can be replaced by the hubs from $L_{in}(a)\cup L_{out}(b)$. Then partial directed BFS rooted at each affected hub can be conducted from $a$ or $b$ depending on whether it is from $L_{in}(a)$ or $L_{out}(b)$. Specifically, hubs from $L_{in}(a)$ will perform BFS from $b$ in forward direction and they will generate or update in-labels (labels in $(L_{in})$) for other vertices. And vice versa for those hubs from $L_{out}(b)$ which affect out-labels for others. For decremental updates, assume edge $(a,b)$ is deleted. The affected vertices sets are defined as follows. The notations are the same as for undirected graphs, but with directions. $SR_a=\{v|\texttt{sd}(v,a)+1=\texttt{sd}(v,b)\land ((v\in L_{in}(a)\land v\in L_{in}(b)) \lor \texttt{spc}(v,a)=\texttt{spc}(v,b))\}$, and $SR_b=\{v|\texttt{sd}(b,v)+1=\texttt{sd}(a,v)\land ((v\in L_{out}(a)\land v\in L_{out}(b)) \lor \texttt{spc}(a,v)=\texttt{spc}(b,v))\}$. Similarly, $R_a=\{v|\texttt{sd}(v,a)+1=\texttt{sd}(v,b)\land v\notin SR_a\}$, and $R_b=\{v|\texttt{sd}(b,v)+1=\texttt{sd}(a,v)\land v\notin SR_b\}$. The BFS for updating is performed from vertices in $SR_a\cup SR_b$ and follows the similar steps as for undirected graph, but with directions.

\subsection{Extension To Weighted Graphs}
For weighted graphs, the labels store the sum of weights along the shortest paths instead of the number of hops. Dijkstra's algorithm replaces BFS for index construction, and a priority queue is used instead of a FIFO queue to track visited vertices. Incremental updates to the index follow a similar approach as \textsc{IncSPC}. When an edge $(a,b)$ with weight $w_{ab}$ is inserted, the affected hubs come from $L(a) \cup L(b)$. Starting from $b$, a partial Dijkstra-like execution is performed with an initial distance of $d_{hb} + w_{ab}$ and initial path counting of $c_{hb}$, where $(h,d_{hb},c_{hb}) \in L(a)$. New compliant shortest paths are searched for and updated if necessary. Decreasing the weight of an existing edge $(a,b)$ is also treated as an incremental update. If the weight decreases from $w_{ab}$ to $w'_{ab}$, the update process is identical to the previous method, using an initial distance of $d_{hb} + w'_{ab}$. For edge deletion or weight increase cases, the conditions for the $SR$ and $R$ sets remain applicable. When an edge $(a,b)$ with weight $w_{ab}$ is deleted or its weight decreases from $w_{ab}$ to $w'_{ab}$, the distance constraint for affected vertices is based on weight rather than the number of hops, i.e., $|\texttt{sd}(v,a)-\texttt{sd}(v,b)|=w_{ab}$. The main difference when applying Algorithm~\ref{alg:srr_search} and Algorithm~\ref{alg:dec_update} to find affected vertices and search for updates is the use of a Dijkstra-like search.

\end{document}

%% file: 1_introduction.tex
\section{Introduction}
Given two vertices $s$ and $t$ in a graph $G$, the shortest path between $s$ and $t$ is the path between them with the shortest length (i.e., the minimum number of edges). As one of the most fundamental notions in graph analytics, it is utilised in a variety of applications, such as group betweenness evaluation~\cite{brandes2001faster,puzis2007fast}, influential community search~\cite{li2017most}, link analysis~\cite{yen2008family}, route planning~\cite{abraham2012hierarchical,abraham2011hub}, and keyword search~\cite{he2007blinks,jiang2015exact}. The length of the shortest paths between $s$ and $t$, i.e., their distance, can be used to measure the correlation between the two vertices. In social networks, for instance, the distance between users can be employed as a parameter in the ranking mechanism of search results, so that the most relevant results are displayed first. Moreover, in collaboration networks, distances between two authors provide a quantitative metric for their collaboration, with Erd\H{o}s number serving as a notable example.

\begin{figure}[t]
  \centering
  \includegraphics[scale=0.5]{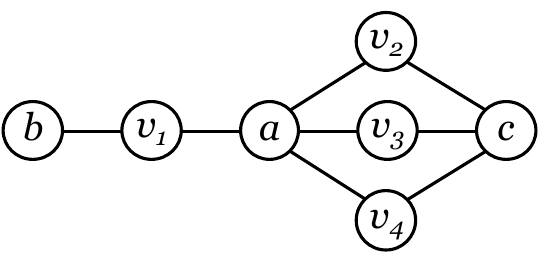}
  \caption{Example Graph $H$}
  \label{fig:app}

\end{figure}

Measuring the correlation based merely on distance is insufficient \rev{in some instances}. The diameter, i.e., the length of the longest shortest path in a graph, is generally small for many networks, reflecting the small-world phenomenon in real-life networks. The distance between a large number of pairs of vertices is distributed within a limited range. Therefore, using the shortest distance alone is \rev{sometimes} not adequate in reflecting the correlations between vertices. Consider the graph $H$ as shown in Figure~\ref{fig:app}, which represents a toy example for a social network where the vertices and edges denote users and friendship, respectively. In $H$, both user $b$ and $c$ are at a distance of $2$ from user $a$. Consequently, based merely on the distances, $b$ and $c$ are equally related in friendship with $a$. However, the result does not match the intuition, as there are more shortest paths between $a$ and $c$. It indicates that $a$ and $c$ are more likely to be friends since they have more common friends. User $c$ will be ranked first when recommending friends for $a$. Similarly, graph $H$ can also simulate a small collaboration network. Vertices and edges are authors and co-authorship in research papers, respectively. Based on the same observation of the shortest path counting, author $c$ is more likely to work in the same field as author $a$. 

The computation of group betweenness exemplifies another application of the shortest path counting problem ~\cite{zhang2020hub}. Group betweenness estimates the importance of a vertex set $C$ to a graph $G$ with vertex set $V$ \cite{puzis2007fast}. Specifically, let $SP(s, t)$ and $\delta_{s,t}$ represent the set of all shortest paths and the number of shortest paths between any vertex pair $(s,t)$, respectively. The number of the paths in $SP(s, t)$ via $C$ is denoted as $\delta_{s,t}(C)$. Group Betweenness of $C$, indicated by \"B$(C)$, is defined as \"B$(C) = \sum_{s, t}\delta_{s,t}(C) / \delta_{s,t}$ where $s,t\in V\setminus C \wedge s\not=t$. As shown in~\cite{puzis2007fast}, shortest path counting between two vertices is the essential building block in the computation of betweenness centrality. 

Shortest path counting over conventional (i.e., static) graphs is first investigated in ~\cite{zhang2020hub}. A straightforward \rev{way to compute the number of shortest paths between two vertices} is to process a breadth-first search (BFS) rooted at one of the vertices, e.g., $s$, and terminated at the other vertex, e.g., $t$. During the graph traversal in a breadth-first fashion, we keep track of and update the shortest distance $D[v]$ from $s$ to each visited vertex $v$, as well as the counting of shortest paths $C[v]$ from $s$ to $v$. Initially, we set $D[s]=0$ and $C[s]=1$. When a vertex $w$ is visited for the first time through its neighbor $v$, we can derive $D[w]=D[v]+1$ and $C[w]=C[v]$. If $w$ has been visited and $D[w]=D[v]+1$, it means new shortest paths from $s$ to $w$ are discovered so that $C[w]$ is updated to $C[w]+C[v]$. After the BFS terminated at $t$, $C[t]$ is the shortest path counting as required. However, real-world networks can be large, leading to lengthy and unsteady execution times for BFS. To address this, 2-hop labeling was developed for efficient shortest path counting in static graphs~\cite{zhang2020hub}. It involves pre-computing an index $L$ as a hub labeling. Each vertex $v$ has its own label set $L(v)$ containing a limited number of hub vertices and other additional information. \rev{This information includes the shortest distance from the hubs to $v$ and a portion of the shortest path counting between them.} 
The shortest path counting between two vertices, e.g., $s$ and $t$, can be calculated in linear time by merely scanning $L(s)$ and $L(t)$. However, in the applications above, the underlying graphs can be highly dynamic and the shortest path counting query can be issued frequently to support continuous tasks such as recommendation or computation. For instance, in a social network, new users are registered and follow new friends, captured by the insertion of vertices and edges, respectively. Additionally, edge deletions can indicate unfollowing operations between users. In these cases, the pre-computed 2-hop labeling index (SPC-Index) in ~\cite{zhang2020hub} can no longer answer the shortest path counting query correctly with the changes of the graph.  

\textbf{Challenges and Contributions:}~Reconstructing the index after each change or after a batch of changes is a straightforward solution to the aforementioned issue. Due to the massive scale of real-world networks, indexing might take hundreds or thousands of minutes, resulting in reconstruction impractical. In the last decade, dynamic hub labeling for shortest distance has been intensively explored, and the decremental update is regarded as extremely challenging due to its inflexible complexity. Due to the additional information that is required to be updated for the shortest path counting, it is difficult to maintain the SPC-Index in an acceptable running time for both incremental and decremental updates. \rev{The existing update algorithms for shortest distance labeling are based on the strategy of partial index update/reconstruction, which has proven to be effective. However, these algorithms cannot be trivially applied to the SPC-Index due to either a large number of redundant computations or insufficient updates. Further discussion on this topic is available in the existing works section. As a result, efficiently searching and updating the affected parts of the SPC-Index following graph changes is a challenging task.}
To cope with these issues, we carefully design the update algorithms, which only focus on the affected vertices. Following that, the search algorithm from these affected vertices is delicately designed to reduce the traversal cost.

We address the aforementioned issues in this paper and present the following contributions.

\noindent
{\em (1) Incremental Update Algorithm \textsc{IncSPC}.} \rev{We study the maintenance of hub labeling for shortest path counting and propose the first efficient incremental update algorithm \textsc{IncSPC} for an added vertex/edge in order to maintain the SPC-Index. Instead of reconstructing everything, our proposed method focuses on discovering emerging shortest paths and updating only the affected portions of the index which lowers the time cost.}

\noindent
{\em (2) Decremental Update Algorithm \textsc{DecSPC}.} \rev{
We propose \textsc{DecSPC}, which is the first decremental update algorithm designed for the SPC-Index that is capable of handling vertex/edge deletion cases. Our approach is based on a novel search-update procedure, and we define innovative sets of affected vertices to enable the detection of a limited number of affected hubs, thereby reducing the overall complexity associated with decremental updates.}

\noindent
{\em (3) Effectiveness and Efficiency.} We conduct extensive experiments on ten graphs to prove that our algorithms are efficient and effective. The results show that our algorithms are up to four orders of magnitude faster than the reconstruction. Moreover, the query time is about two orders of magnitude faster than the baselines.

\textbf{Roadmap.} The rest of the paper is organized as follows. Section II introduces some preliminaries. Section III investigates the index maintenance for edge insertion and deletion, followed by empirical studies in Section IV. Section V surveys important related work. Section VI concludes the paper.

The proofs, application examples, and extensions can be found in Appendix.

%% file: 2_preliminaries.tex
\section{Preliminaries}

\subsection{Preliminaries}

$G=(V, E)$ denotes an undirected unweighted graph with vertex set $V$ and edge set $E$. Let $n$ and $m$ represent the number of vertices and edges, respectively. An undirected edge $(v,u)\equiv (u,v)\in E$ connects two vertices $v$ and $u\in V$. The set of all $v$'s neighbors is defined as \texttt{nbr}$(v)$. The degree of a vertex $v$ is the number of its connected edges and is defined as \texttt{deg}$(v)$. \rev{A shortest path between two vertices $s$ and $t$, denoted by \texttt{sp}$(s, t)$, is a path with the smallest number of edges connecting them.} We use \texttt{sd}$(s, t)$ to denote the shortest distance from $s$ to $t$, i.e., the length of \texttt{sp}$(s,t)$. The set of all shortest paths from $s$ to $t$ is denoted by \textit{SP}$(s, t)$. Shortest path counting from $s$ to $t$ is defined as \texttt{spc}$(s,t)$, which is the number of shortest paths from $s$ to $t$. 
\texttt{spc}$(\Hat{s},t)$ is the number of shortest paths between $s$ and $t$ which have $s$ as the highest-ranked vertex, \rev{and therefore} $\texttt{spc}(s,t)\geq\texttt{spc}(\Hat{s},t)$. Similarly, $sp(\Hat{s},t)$ and $SP(\Hat{s},t)$ denote one and all shortest path(s) between $s$ and $t$ with $s$ as the highest-ranked vertex, respectively. Table~\ref{tab:notation} summarizes the key notations used throughout this paper.

\begin{table}[t]
\centering
\caption{Notations}
\vspace{-1mm}
  \begin{tabular}{|l|l|}
    \hline
    \cellcolor{gray!25}\textbf{Notations} & \cellcolor{gray!25}\textbf{Description}\\
    \hline
     $G = (V, E)$ & \makecell[l] {an undirected and unweighted graph}\\
    \hline
    $G_{i}$ & the graph after i updates\\
    \hline
    $n$, $m$ & \makecell[l] {the number of vertices and edges}\\
    \hline
    \texttt{deg}$(v)$ & the degree of vertex $v$\\
    \hline
    \texttt{nbr}$(v)$ & the neighbors of vertex $v$\\
    \hline
    \texttt{sd}$(s, t)$ & the shortest distance from $s$ to $t$\\
    \hline
    \texttt{sd}$_{i}(s, t)$ & the shortest distance from $s$ to $t$ in $G_{i}$\\
    \hline
    $\texttt{spc}(s, t)$ & the \# of shortest paths from $s$ to $t$\\
    \hline
    $\texttt{spc}_{i}(s, t)$ & \makecell[l] {the \# of shortest paths from $s$ to $t$ in $G_{i}$}\\
    \hline
    $\texttt{spc}(\Hat{s}, t)$ & \makecell[l] {the \# of shortest paths from $s$ to $t$ where \\ $s$ has the highest rank}\\
    \hline
\end{tabular}
\label{tab:notation}
\end{table}

\subsection{Hub Labeling for Shortest Path Counting}

The shortest path counting query \texttt{SPC}$(s, t)$ is to determine the number of shortest paths between $s$ and $t$, i.e., \texttt{spc}$(s,t)$. \cite{zhang2020hub} extended Pruned Landmark Labeling (PLL), the state-of-the-art to calculate shortest distances in real-time on large-scale graphs, and proposed a hub-labeling algorithm to answer \texttt{SPC}$(\cdot,\cdot)$ queries on static graphs. \rev{We say the labeling obeys the cover constraint, Exact Shortest Paths Covering (ESPC), if \texttt{spc}$(s,t)$ can be calculated for any $s$ and $t$ with Equation \eqref{con:spc1} and \eqref{con:spc2}. In other words, it encodes not only the shortest distance between any two vertices but also ensures that the count of such shortest paths is accurate.} In the labeling, assume $\leq$ is the total order over $V$. We say $v$ has a higher rank than $u$ if $v \leq u$. \rev{Each vertex $v$ owns a label set $L(v)$, which is a set of triples of the form $\{(h,\texttt{sd}(h,v),\sigma_{h,v})\}$ where $\sigma_{h,v}=\texttt{spc}(\Hat{h},v)$.} In case of \rev{$\sigma_{h,v}=\texttt{spc}(h,v)$}, $(h,\texttt{sd}(h,v),\sigma_{h,v})$ is a canonical label. Otherwise, $(h,\texttt{sd}(h,v),\sigma_{h,v})$ is a non-canonical label. We use $h\in L(v)$ to represent \rev{that} $h$ is a hub in $L(v)$. With the SPC-Index, \texttt{spc}$(s,t)$ can be computed based on the following equation.

\begin{equation}
    H = \{h | \mathop{\arg\min}\limits_{h \in L(s) \cap L(t)} (\texttt{sd}(h,s) + \texttt{sd}(h,t))\} 
    \label{con:spc1}
\end{equation}

\begin{equation}
    \texttt{spc}(s,t) = \sum_{h\in H} \sigma(h,s) \cdot \sigma(h,t) 
    \label{con:spc2}
\end{equation}

Equation \eqref{con:spc1} finds the common hubs $H$ in $L(s)$ and $L(t)$ which lie on the shortest paths between $s$ and $t$. Equation \eqref{con:spc2} computes \texttt{spc}$(s,t)$ based on $H$. The query evaluation can be implemented by Algorithm~\ref{alg:spc_query}. For each common hub $h$ (line 2), if the distance between $s$ and $t$ via $h$ is smaller than the temporary distance $d$, reset both temporary distance and counting (lines 3-5). If the distance via $h$ equals the temporary distance, accumulate the counting (lines 6-7). Note that if $s$ and $t$ are disconnected, there is no common hub in $L(s)$ and $L(t)$ so that $d$ and $c$ will remain $\infty$ and 0, respectively.

\begin{figure} [t]
    \centering
    \includegraphics[scale=0.6]{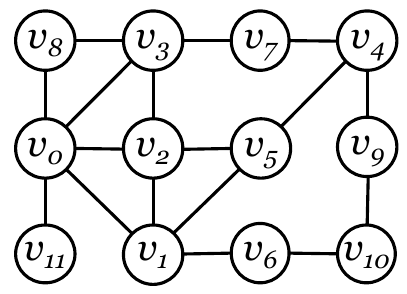}
    \caption{Example Graph $G$.}
    \label{fig:ori_g}
\end{figure}

Figure \ref{fig:ori_g} illustrates an undirected graph with 12 vertices and \autoref{tab:spc_index_fig1} presents its SPC-index. Assume the vertices are assigned with the following ordering: $v_0\leq v_1\leq v_2\leq v_3\leq v_4\leq v_5\leq v_6\leq v_7\leq v_8\leq v_9\leq v_{10}\leq v_{11}$.

\begin{table}[t]
\small
\centering
\caption{SPC-Index of Graph $G$ in Figure~\ref{fig:ori_g}}
\begin{tabular}{|l|l|l|}
    \hline
    \cellcolor{gray!25}\textbf{$v$} & \cellcolor{gray!25}\textbf{$L(v)$}\\
    \hline
    $v_0$ & $(v_0,0,1)$ \\
    \hline
    $v_1$ & $(v_0,1,1)$ $(v_1,0,1)$ \\
    \hline
    $v_2$ & $(v_0,1,1)$ $(v_1,1,1)$ $(v_2,0,1)$ \\
    \hline
    $v_3$ & $(v_0,1,1)$ $(v_1,2,1)$ $(v_2,1,1)$ $(v_3,0,1)$ \\
    \hline
    $v_4$ & $(v_0,3,3)$ $(v_1,2,1)$ $(v_2,2,1)$ $(v_3,2,1)$ $(v_4,0,1)$ \\
    \hline
    $v_5$ & $(v_0,2,2)$ $(v_1,1,1)$ $(v_2,1,1)$ $(v_4,1,1)$ $(v_5,0,1)$ \\
    \hline
    $v_6$ & $(v_0,2,1)$ $(v_1,1,1)$ $(v_4,3,1)$ $(v_6,0,1)$ \\
    \hline
    $v_7$ & $(v_0,2,1)$ $(v_1,3,2)$ $(v_2,2,1)$ $(v_3,1,1)$ $(v_4,1,1)$ $(v_7,0,1)$ \\
    \hline
    $v_8$ & $(v_0,1,1)$ $(v_2,2,1)$ $(v_3,1,1)$ $(v_8,0,1)$ \\
    \hline
    $v_9$ & \makecell[l]{$(v_0,4,4)$ $(v_1,3,2)$ $(v_2,3,1)$ $(v_3,3,1)$ $(v_4,1,1)$ $(v_6,2,1)$ \\$(v_9,0,1)$} \\
    \hline
    $v_{10}$ & \makecell[l]{$(v_0,3,1)$ $(v_1,2,1)$ $(v_3,4,1)$ $(v_4,2,1)$ $(v_6,1,1)$ $(v_9,1,1)$ \\$(v_{10},0,1)$} \\
    \hline
    $v_{11}$ & $(v_0,1,1)$ $(v_{11},0,1)$ \\
  \hline
\end{tabular}
\label{tab:spc_index_fig1}
\end{table}

\begin{example}
Considering the query \texttt{SPC}$(v_4, v_6)$. By scanning their common hubs $\{v_0, v_1, v_4\}$, we determine that $H=\{v_1, v_4\}$ and \texttt{sd}$(v_4, v_6)=\texttt{sd}(v_1, v_4)+\texttt{sd}(v_1,v_6) = \texttt{sd}(v_4, v_4) + \texttt{sd}(v_4, v_6) = 3$. Thus, \texttt{spc}$(v_4, v_6) = \sigma_{v_1,v_4}\cdot \sigma_{v_1,v_6} + \sigma_{v_4,v_4}\cdot \sigma_{v_4,v_6} = 1\times1+1\times1=2$.
\end{example}

\begin{example}
In $L(v_5)$, label entry $(v_0, 2, 2)$ is canonical because $v_0$ has the highest rank along all the shortest paths between $v_0$ and $v_5$, i.e., \texttt{spc}$(\Hat{v_0},v_5)=$\texttt{spc}$(v_0, v_5)=2$. In $L(v_8)$, $(v_2, 2, 1)$ is a non-canonical label, because \texttt{spc}$(\Hat{v_2},v_8)=1<$\texttt{spc}$(v_2,v_8)$ and the corresponding path is $v_2\rightarrow v_3\rightarrow v_8$.
\end{example}

\cite{zhang2020hub} also proposed a hub pushing algorithm namely \texttt{HP-SPC} to construct the SPC-Index. Given a graph $G$ and total order $\leq$. Each vertex $v$, in descending order, conducts a breadth-first search in $G_v$, where $G_v$ is the subgraph that only includes the vertices of which rank is not higher than $v$. The shortest distance and the shortest path counting from $v$ are recorded during the traversal. When a vertex $w$ is explored, BFS is pruned if a shorter distance from $v$ to $w$ can be computed with the existing index. Otherwise, a label with $v$ as the hub is generated and inserted into $L(w)$.

\begin{algorithm}[htb]
    $d \leftarrow \infty$; $c \leftarrow 0$\;
    \For{{\bf each} $h \in L(s)\cap L(t)$} {
        \If{\texttt{sd}$(h,s) + \texttt{sd}(h,t) < d$} {
            $d \leftarrow \texttt{sd}(h,s) + \texttt{sd}(h,t)$\;
            $c \leftarrow \sigma_{h,s} \cdot \sigma_{h,t}$\;
        } \ElseIf{$\texttt{sd}(h,s) + \texttt{sd}(h,t) = d$} {
            $c \leftarrow c + \sigma_{h,s} \cdot \sigma_{h,t}$\;
        }
    }
    \textbf{return} $(d,c)$\;
\caption{\textsc{Spc}QUERY($s,t$)}
\label{alg:spc_query}
\end{algorithm}

A well-designed vertex ordering can significantly reduce the index construction time, index size, and query time. Intuitively, vertices with larger degrees are considered to lie on more shortest paths and thus, are ranked higher so that the later searches in \texttt{HP-SPC} can be pruned earlier. The degree-based ordering, which sorts vertices based on descending degrees, is widely used in 2-hop labeling schemas and is adopted in our work as established by \cite{zhang2020hub}.

On static graphs, hub labeling for shortest path counting has been extensively researched~\cite{zhang2020hub}. To be consistent with the very dynamic nature of modern networks, the graphs' topological structure may change over time.
Thus, the dynamic labeling for shortest path counting problem is defined as follows:

{\bf Problem Statement.} Given a graph $G$ and its SPC-Index $L$, we intend to maintain $L$ accordance with the topological modifications, i.e., insertion and deletion of vertices and edges, applied to $G$.

\subsection{Existing Works}
Conventionally, shortest path counting queries are answered by online algorithms like Dijkstra's algorithm or Breadth-First Search (BFS) or bi-directional BFS which conducts searches from both query vertices concurrently. Although these online algorithms are not impacted by graph changes, they still lead to unpredictable and potentially lengthy response times as each query requires a new graph traversal. In contrast, due to the lack of research on dynamic hub labeling for shortest path counting, existing dynamic 2-hop labeling methods primarily concentrate on maintaining shortest distance indexes are summarized as follows.

The 2-hop labeling for shortest distance, e.g., Pruned Landmark Labeling\cite{akiba2013fast}, involves precomputing an index for the entire graph, enabling query evaluation to be performed exclusively through the generated index, thereby obviating the necessity for graph traversal. \rev{The SD-Index, which represents the shortest distance index under PLL, only contains the hubs from canonical labels in SPC-Index and the corresponding shortest distances. As opposed to Example 1, $(v_0,2)$ belongs to $L(v_5)$ in SD-Index, but $v_2$ is no longer a hub of $v_8$. This is because to answer the shortest distance query between any pair of vertices, SD-Index only requires the distance information of one hub lying on any one of the shortest paths between them. In contrast, for SPC queries, all the shortest paths between the vertices should be counted. Thus, both canonical and non-canonical labels with counting information are required. We apply a similar strategy of partial update used in existing works for updating the SD-Index to update the SPC-Index due to its efficiency and effectiveness. Although both SPC-Index and SD-Index comply with the PLL schema, existing update algorithms for SD-Index cannot be trivially extended to SPC-Index because the additional non-canonical labels and counting information complicate the design.}

For the maintenance of SD-Index, \cite{akiba2014dynamic} proposed an incremental update algorithm for vertex/edge insertion. To reduce the search space during the update, the algorithm performs pruned BFS that is rooted at the hubs of endpoints of the inserted edge, in order to discover new shortest paths. \rev{However, their algorithm lacks the capability to update the SPC-Index when inserting new edges due to the inadequate pruning condition that fails to detect the presence of new shortest paths with the same length as the pre-existing ones. Furthermore, their approach only records the shortest distances during the BFS procedures, but not the number of shortest paths.}

\cite{qin2017efficient,d2019fully} studied the problem of updating the SD-Index for vertex/edge deletion. Firstly, they define affected vertices based on the deleted edge. The labels of these affected vertices or the labels with them as hubs may be invalid. Therefore, all their related labels are then removed. Afterward, the portion of the index that is missing is reconstructed. \rev{These algorithms cannot be trivially adapted to the maintenance of the SPC-Index. Assume edge ($a,b$) is deleted from $G_i$ and the resulting graph is $G_{i+1}$. If we apply the definition of affected vertices proposed in \cite{d2019fully} to find the affected vertices for SPC-Index. It will result in a collection of all vertices $v$ such that $sd_i(v,a)+1=sd_i(v,b)$ or $sd_i(v,b)+1=sd_i(v,a)$ because SPC-Index covers all the shortest paths between any pair of vertices. However, this set of affected vertices doesn't provide any reduction and includes a large number of unnecessary vertices such that labels that use them as hubs don't need to be updated.}

\rev{In \cite{qin2017efficient}, the authors propose real affected vertices and possible affected vertices to identify invalid labels and missing labels. According to their definition, a vertex $v$ belongs to the set of real affected vertices $RA(a)$ if $sd_i(v,a)+1=sd_i(v,b) \land sd_i(v,a)+1\neq sd_{i+1}(v,b)$ or $RA(b)$ if $sd_i(v,b)+1=sd_i(v,a)\land sd_i(v,b)+1\neq sd_{i+1}(v,a)$. For any given $s\in RA(a)$ and $t\in RA(b)$ or vice versa, $(s, sd(s,t))$ is an invalid label if $(s, sd(s,t))\in L(t)$. However, in the case of SPC-Index, deleting an edge may reduce the number of shortest paths between two vertices. but their respective shortest distance could remain unchanged. For instance, assume $(v,d,c)\in L(b)$, and edge $(a,b)$ is subsequently removed. $d$ keeps constant but $c$ decreases. In such a case, the label with hub $v$ in $L(b)$ requires an update, but neither $RA(a)$ nor $RA(b)$ contains $v$.} 

%% file: 3_incremental.tex
\section{DSPC: Dynamic Shortest Path Counting}
The na\"ive method to deal with the changes is to reconstruct the index for each update. Nevertheless, the time cost equals the indexing time, which is too costly. To fulfill this research gap, we propose a \underline{D}ynamic \underline{S}hortest \underline{P}ath \underline{C}ounting algorithm, i.e., DSPC. It comprises of two algorithms to update the SPC-Index in the case of edge insertion and deletion, respectively. For a newly-added isolated vertex $v$, we only need to add an empty label set $L(v)$ to the original index. For the deletion of an existing vertex $v$, it is equivalent to deleting all its connected edges from the graph. The index can be updated by executing a sequence of our decremental algorithm.

\subsection{Incremental Update}
Suppose edge $(a, b)$ is inserted into $G_i$, with SPC-Index $L_i$, and the resulting graph is $G_{i+1}$.
To correctly update $L_{i}$ and avoid building the index from scratch, we define the affected hub set \texttt{AFF} based on $(a, b)$. Labels with $h\in \texttt{AFF}$ as hub may be outdated or should be added. 
The main idea of our incremental update algorithm is to conduct pruned BFSs from affected hubs and update labels during the traversal. 
We first introduce some crucial lemmas and analyze the case of potential label updates due to \texttt{sd} and \texttt{spc} changes caused by edge insertion. Then extend to the maintenance of the SPC-Index.

\begin{lemma} 
    Suppose a new edge $(a, b)$ is inserted into $G_i$, and the resulting graph is $G_{i+1}$. For any two arbitrary vertices $v$ and $u$, $\texttt{sd}_{i+1}(v,u)\leq \texttt{sd}_i(v,u)$.
\label{lemma:label_keep}
\end{lemma}

According to Lemma \ref{lemma:label_keep}, shortest distances will never increase with the insertion of a new edge. Those distance-outdated labels are not necessary to be removed. Because our query algorithm always returns the minimum distance, overestimated distances will be eliminated invariably. The retention of outdated labels will not affect the correctness of queries but can reduce the update time. 

\begin{lemma}
    Suppose a new edge $(a, b)$ is inserted into $G_i$, and the resulting graph is $G_{i+1}$. For any two arbitrary vertices $v$ and $u$, if $\texttt{sd}_i(v,u)\not=\texttt{sd}_{i+1}(v,u)$, then $\texttt{sp}_{i+1}(v,u)$ passes through $(a,b)$ for all $\texttt{sp}_{i+1}(v,u)\in SP_{i+1}(v,u)$.
\end{lemma}

\begin{lemma}
    Suppose a new edge $(a, b)$ is inserted into $G_i$, and the resulting graph is $G_{i+1}$. Assume $w$ is the penultimate vertex in $\texttt{sp}_i(v,u)$ where $v$ and $u\not= a,b$. If $\texttt{sp}_{i}(v,u)$ has changed in $G_{i+1}$, then $\texttt{sd}_{i+1}(v,w)\not=\texttt{sd}_{i}(v,w)$.
\end{lemma}

Assume $v$ is an affected hub and $\texttt{sd}_i(v,a) \leq \texttt{sd}_i(v,b)$. BFS can be conducted from $v$ to search the vertices whose label $(v,\cdot,\cdot)$ should be updated. Let $D[\cdot]$ and $C[\cdot]$ store the BFS tentative \texttt{sd} and \texttt{spc} from $v$. Based on the above lemmas, BFS can be partially started from $b$ with distance $D[b]=\texttt{sd}_i(v,a)+1$ and path counting $C[b]=\texttt{spc}_i(v,a)$. Just like passing through $(a,b)$ from $a$. The BFS will find all vertices $w$ s.t. $\texttt{sd}_{i+1}(v,w)\not=\texttt{sd}_i(v,w)$ if it terminates when distance doesn't change, i.e., $D[v_t]=\texttt{sd}_i(v,v_t)$ where $v_t$ is the current visited vertex. For all $w$, the label $(v,d,c)\in L(w)$ should be renewed with new distance if necessary. For the affected hub with $\texttt{sd}_i(v,b) \leq \texttt{sd}_i(v,a)$, BFS will be conducted from $a$ with $D[a]=\texttt{sd}_i(v,b)+1$ and $C[a]=\texttt{spc}_i(v,b)$. We present the following lemma for further analysis. For simplicity, we use $\overline{\texttt{spc}}^{i}_{i+1}(v,u)$ to indicate the fact between $v$ and $u$ such that $\texttt{spc}_{i+1}(v,u)\not=\texttt{spc}_i(v,u)$ but $\texttt{sd}_{i+1}(v,u)=\texttt{sd}_i(v,u)$.

\begin{figure*}[t]
	\begin{center}
		\subfigure[Update Rooted at Hub $v_0$.]{
			\label{fig:inc_0}
			\centering
			\includegraphics[scale = 0.55]{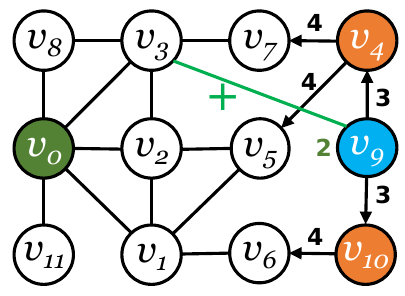} 
		} \hspace{4mm}
		\subfigure[Update Rooted at Hub $v_1$.]{
			\label{fig:inc_1}
			\centering
			\includegraphics[scale = 0.55]{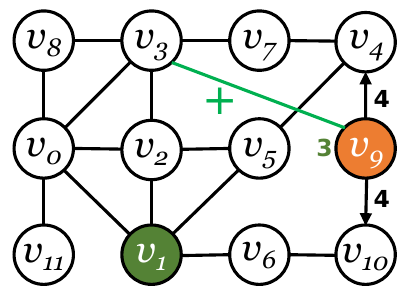}     
		} \hspace{4mm}
        \subfigure[Update Rooted at Hub $v_2$.]{
			\label{fig:inc_2}
			\centering
			\includegraphics[scale = 0.55]{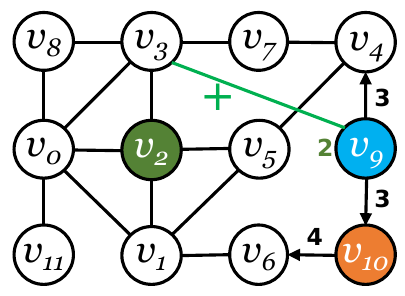}     
		}\vspace{-1mm}
        \subfigure[Update Steps for Affected Hubs $v_0$, $v_1$ and $v_2$]{\footnotesize
            \label{fig:inc_1_table}
            \centering
            \begin{tabular}{c||c|c|c||c|c|c}
              \toprule
              \makecell[l] {\emph{BFS Distance}} & \makecell[l] {\emph{Meeting Vertex}} &
              \makecell[l] {\emph{Rank Pruning}}  & \makecell[l] {\emph{Distance Pruning}} &
              \makecell[l] {\emph{Existing Label}}&
              \makecell[l] {\emph{D[$\cdot$], C[$\cdot$]}}    & \makecell[l] {\emph{Update Operation and New Label}}\\\bottomrule
              \multicolumn{7}{c}{Affected Hub $v_0$, BFS from $v_9$ with D[$v_9$]=2 and C[$v_9$]=1}\\\toprule
                          D=2    & $v_9$  & Pass & Pass ($d_L$=4>D)   & ($v_0,4,4$) & 2,1 & Renew d and c ($v_0,2,1$)  \\\cline{1-7}
              \multirow{2}*{D=3} & $v_4$  & Pass & Pass (3=D)   & ($v_0,3,3$) & 3,1 & Renew c ($v_1,3,4$)  \\\cline{2-7}
              ~                  &$v_{10}$& Pass & Pass (3=D)   & ($v_0,3,1$) & 3,1 & Renew c ($v_1,3,2$)  \\\hline
                          D=4    & $v_5$ $v_7$ $v_6$ & Pass & Pruned (2<D) &      -      &  -  & -                  \\\bottomrule
              \multicolumn{7}{c}{Affected Hub $v_1$, BFS from $v_9$ with D[$v_9$]=3 and C[$v_9$]=1}\\\toprule
                         D=3     & $v_9$  & Pass & Pass (3=D)   & ($v_1,3,2$) & 3,1 & Renew c ($v_1,3,3$)  \\\hline
                         D=4     & $v_4$ $v_{10}$ & Pass & Pruned (2<D) &      -      &  -  & -                  \\\bottomrule
              \multicolumn{7}{c}{Affected Hub $v_2$, BFS from $v_9$ with D[$v_9$]=2 and C[$v_9$]=1}\\\toprule
                         D=2     & $v_9$  & Pass & Pass (3>D)   & ($v_2,3,1$) & 2,1 & Renew d and c ($v_2,2,1$)  \\\hline
              \multirow{2}*{D=3} & $v_4$  & Pass & Pruned (2<D) &      -      &  -  & -                  \\\cline{2-7}
              ~                  &$v_{10}$& Pass & Pass (3=D)   &  No Label   & 3,1 & Insert ($v_2,3,1$) \\\hline
                         D=4     & $v_6$  & Pass & Pruned (2<D) &      -      &  -  & -                  \\\bottomrule
            \end{tabular}
        }
	\end{center}
    \vspace{-4mm}
	\caption{Example of Incremental Update. Insert Edge $(v_3, v_9)$ to $G$.} \label{fig:inc_example}
\end{figure*}

\begin{lemma}
    Suppose a new edge $(a, b)$ is inserted into $G_i$, and the resulting graph is $G_{i+1}$. For any two vertices $v$ and $u\not=a,b$. Assume $\overline{\texttt{spc}}^{i}_{i+1}(v,u)$ holds and there is a new shortest path $\texttt{sp}_{i+1}(v,u)$ that was absent in $G_i$. Then either $\overline{\texttt{spc}}^{i}_{i+1}(v,w)$ holds or $\texttt{sd}_{i+1}(v,w)\not=\texttt{sd}_i(v,w)$ where $w$ is the penultimate vertex in $\texttt{sp}_{i+1}(v,u)$. 
\label{lemma:cnt_eq}
\end{lemma}

By referring to Lemma \ref{lemma:cnt_eq}, the pruning condition of the BFS should be relaxed to $\texttt{sd}_i(v,v_t)<D[v_t]$ so that all vertices $w$ can be found. Then, $\overline{\texttt{spc}}^{i}_{i+1}(v,w)$ holds. In this case, label $(v,\cdot,\cdot)\in L(w)$ should be renewed with new shortest path counting if necessary. By performing the above pruned BFS, all vertices $w$ such that $\texttt{sd}_{i+1}(v,w)\not=\texttt{sd}_i(v,w)$ or holding $\overline{\texttt{spc}}^{i}_{i+1}(v,w)$ will be reached s.t. all potential label updates will be revealed.

\begin{example}
    Figure~\ref{fig:inc_0} shows an example of inserting edge $(v_3,v_9)$ to the example graph $G$. Assume $v_0$ in green is the affected hub currently processed. The blue nodes represent vertices whose shortest distance to $v_0$ has changed, and the orange nodes represent vertices where only the shortest path counting to $v_0$ has changed. Because $\texttt{sd}_i(v_0,v_3)<\texttt{sd}_i(v_0,v_9)$, the pruned BFS can start from $v_9$ to search for updates with $D[v_9]=\texttt{sd}_i(v_0,v_3)+1=2$ and $C[v_9]=\texttt{spc}_i(v_0,v_3)=1$.
    \label{ex_insert}
\end{example}

To extend the above idea in maintaining SPC-Index, some prerequisites should be applied:
\begin{itemize}
    \item Assume the currently affected hub is $v$ and $\texttt{sd}(v,a)\leq\texttt{sd}(v,b)$. The pruned BFS starting distance and path counting should use the corresponding element from $L_i(a)$ because the label records $\texttt{spc}(\Hat{v},\cdot)$ instead of $\texttt{spc}(v,\cdot)$. Suppose $(v,d,c)\in L_i(a)$. BFS should start from $b$ with $D[b]=d+1$ and $C[b]=c$.
    \item Ranking pruning should be strictly observed to keep ESPC. Suppose the pruned BFS is conducted rooted at $v$, it should terminate when meeting $u$ s.t. $u\leq v$. 
    \item Use the distance calculated with the up-to-date index to compare with BFS tentative distance because some shortest distances may change during the update. 
\end{itemize}

\rev{Finally, $\texttt{AFF}$ is defined as $\{h|h \in L_i(a)\cup L_i(b)\}$, the roots of the pruned BFSs. They are sufficient to cover all potential updates. Because if $h\notin L_i(a)\cup L_i(b)$, $a$ and $b$ are either pruned before or unreachable from $h$. For the example in Figure~\ref{fig:inc_example}, \texttt{AFF}$=\{v_0, v_1, v_2, v_3, v_4, v_6, v_9\}$. Labels with other vertices as hub will not be impacted, hence they are not in \texttt{AFF}. For instance, $v_8\notin\texttt{AFF}$ even though \texttt{sd}$(v_8,v_9)$ decreases. This is because $v_8\notin L(v_3)\cup L(v_9)$ indicates that there was no \texttt{sp}$(\Hat{v_8},v_3)$ or \texttt{sp}$(\Hat{v_8},v_9)$ before the edge insertion. As a result, no new \texttt{sp}$(\Hat{v_8},\cdot)$ will pass through the new edge $(v_3, v_9)$. New shortest paths between $v_8$ and other vertices can be covered by other hubs.}
\vspace{-2mm}

\begin{algorithm}[htb]
    $G_{i+1} \leftarrow G_i \oplus (a,b)$\;
    $\texttt{AFF} \leftarrow \{h|h \in L_i(a)\cup L_i(b)\}$\;
    \For{{\bf each} $h\in \texttt{AFF}$ {\rm in descending order}}{
        \If{$h \in L_i(a)$ {\bf and} $h \leq b$}{
            {\textsc{Inc}UPDATE}($h, a, b$)\;
        }
        \If{$h \in L_i(b)$ {\bf and} $h \leq a$}{
            {\textsc{Inc}UPDATE}($h, b, a$)\;
        }
    }
\caption{\textsc{Inc}SPC($G_i, L_i, a,b$)}
\label{alg:spc_inc}
\end{algorithm}


\begin{algorithm}[htb]
    \For{{\bf each} $u\in V$} {
        $D[u] \leftarrow \infty;$ $C[u] \leftarrow 0$\;
    }
    $(h,d,c) \leftarrow$ label with hub $h$ from $L(v_a)$\;
    $D[v_b] \leftarrow d+1;$ $C[v_b] \leftarrow c$\;
    Queue $Q \leftarrow $ $\emptyset$; $Q.{\rm enqueue}(v_b)$\;
    \While{$Q$ {\rm $\neq \emptyset$}} {
        $v \leftarrow Q.{\rm dequeue}()$\;
        $d_L, c_L \leftarrow$ \textsc{Spc}QUERY($h,v$)\;
        {\bf if} $d_L < D[v]$ {\bf then continue}\;
        \If{$(h,d_i,c_i) \in L(v)$}{
            $d \leftarrow D[v]$; $c \leftarrow C[v]$\;
            \If{$d = d_i$} {
                $c \leftarrow c + c_i$\;
            }
            Replace $(h,d_i,c_i)$ with $(h,d,c)$\;
        } \Else{
            Insert $(h,d,c)$ to $L(v)$\;
        }
        \For{{\bf each} $w \in \texttt{nbr}(v)$} {
            \If{$D[w] =\infty$ {\bf and} $h \leq w$}{
                $D[w] \leftarrow D[v] + 1$; $C[w] \leftarrow C[v]$\;
                $Q.{\rm enqueue}(w)$\;
            } \ElseIf {$D[w]  = D[v] + 1$} {
                $C[w] \leftarrow C[w] + C[v]$\;
            }
        }
    }
\caption{\textsc{Inc}UPDATE($h, v_a, v_b$)}
\label{alg:inc_upd}
\end{algorithm}

\rev{Algorithm \ref{alg:spc_inc} describes the incremental update algorithm. Firstly, the edge $(a,b)$ is inserted into $G_i$ to form the updated graph $G_{i+1}$ (line 1). Then, the affected hub set $\texttt{AFF}$ is obtained from $L_i(a)$ and $L_i(b)$ (line 2). For each affected hub $h$, the pruned BFS is executed starting from $a$ or $b$ based on whether $h\in L_i(a)$ or $h\in L_i(b)$, respectively (lines 3-7). Algorithm \ref{alg:inc_upd} describes the procedure of the pruned BFS rooted at $h$. 
The arrays $D[\cdot]$ and $C[\cdot]$ store BFS tentative distance and shortest path counting, respectively, and they are initialized in lines 1 to 4. The queue $Q$ used to store visited vertices is initialized in line 5. The pruned BFS rooted at $h$ is executed from lines 6 to 22. Assume $v$ is currently visited and pop from $Q$ (line 7), the distance between $h$ and $v$ calculated with the current index is $d_L$, then BFS is pruned when $d_L<D[v]$ (lines 8-9). This is because $L(\cdot)$ already covers $SP(h,v)$, and the paths found by the current BFS paths are not the shortest. Otherwise, renew the label of $v$ if $h\in L(v)$ or insert a new label to $L(v)$ if $h\notin L(v)$ (lines 10-16). BFS continues to search for new shortest paths (lines 17-22).}

\begin{example}
    \rev{Followed with Example~\ref{ex_insert}. Figure~\ref{fig:inc_0}-(c) illustrate the pruned BFSs of the incremental update rooted at affected hubs $v_0$, $v_1$, and $v_2$, respectively. The number in green next to $v_9$ is the starting BFS distance which is $D[v_b]$ in line 4 of Algorithm~\ref{alg:inc_upd}. The number on each edge is the BFS distance which is $D[v]$. The table in Figure~\ref{fig:inc_1_table} shows each step of the pruned BFSs.}
    
    \rev{First, we have \texttt{AFF}$=\{v_0, v_1, v_2, v_3, v_4, v_6, v_9\}$. In Figure~\ref{fig:inc_0}, current affected hub is $v_0$ and BFS starts from $v_9$ with $D[v_9]=2$ and $C[v_9]=1$. Because $v_0\leq v_9$ and $d_L=4>D[v_9]$, it means new \texttt{sp}($\Hat{v_0},v_9$) are discovered and $v_9$'s label requires update. As the distance of the existing label ($v_0,4,4$) in $L(v_9)$ is outdated, the label is replaced by $(v_0,2,1)$. In next step, labels in $L(v_4)$ and $L(v_{10})$ are updated by increasing the counting. BFS is pruned at $v_5$, $v_6$ and $v_7$ because $d_L<D[\cdot]$. In Figure~\ref{fig:inc_2}, current affected hub is $v_2$, an update is required when meeting $v_{10}$ but $v_2\notin L(v_{10})$. In this case, a new label is created for $v_{10}$. The update steps for other affected hubs are omitted.}
\end{example}

\begin{theorem} [Correctness]
\vspace{-1mm}
    Given $G_{i+1} = G_i \oplus (a,b)$, $L_i$ is the SPC-Index of $G_i$ and $L_{i+1}$ is the updated SPC-Index by Algorithm \ref{alg:spc_inc}. $L_{i+1}$ obeys ESPC regarding $G_{i+1}$.
    \label{theorem:inc_correct}
\end{theorem}

\begin{theorem} [Time Complexity]
    Assume average label size $|L(v)|=O(l)$ for any vertex $v$ and the pruned BFS visits $O(k)$ vertices for each affected hub. The time complexity of Algorithm \ref{alg:spc_inc} is $O(kl^2)$.
    \label{theorem:inc_time}
\end{theorem}

%% file: 4_decremental.tex

\subsection{Decremental Update}
Different from edge insertion, deleting an edge may increase the shortest distances. Outdated label entries, e.g., their distance elements are smaller than the actual distances, will underestimate the distance part of the queries. For this reason, detecting and updating these label entries is unavoidable. 

Assume edge $(a,b)$ is deleted from $G_i$ and the resulting graph is $G_{i+1}$. We still follow the partial update strategy for the decremental updates. But unlike incremental updates, the BFS cannot start from one of the endpoints of the removed edge because new shortest paths may not overlap the former sub-path before the deleted edge. In addition, affected hubs are not only the ones from $L_i(a)\cup L_i(b)$. 

\begin{figure} [thb]
  \centering
  \includegraphics[width=\linewidth]{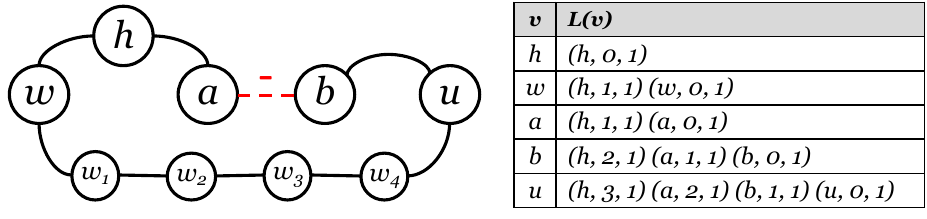}
  \vspace{-3mm}
  \caption{Delete $(a,b)$ From the Toy Graph.}
  \label{fig:insert_example}
\end{figure}

\begin{example}
    Given a toy graph in \autoref{fig:insert_example}. Assume $h\leq w \leq a \leq b \leq u\leq w_1\leq w_2\leq w_3\leq w_4$ and the index in the right shows the labels of $h$, $w$, $a$, $b$, and $u$ before the deletion of edge $(a,b)$. We can see that $h$ is the hub of both $a$ and $b$ but $w$ is neither. After deleting $(a,b)$, $(h,3,1)$ in $L(u)$ should be updated to $(h,6,1)$ and the corresponding path does not pass $(h,a)$ anymore. Besides, $(w,5,1)$ should be added into $L(u)$ despite $w$ was not the hub of $a$ or $b$.
\end{example}

\rev{For any vertex $v$, if there exists a shortest path from it and through $(a, b)$, then we have  $|\texttt{sd}_i(v,a)-\texttt{sd}_i(v,b)|=1$. Intuitively, all such vertices may be affected by the deletion of $(a,b)$. However, updating the index by conducting a BFS from each of these vertices can be computationally expensive. To limit the number of affected vertices and reduce the number of BFS required, we can categorize the affected vertices into two types. }
\rev{
\begin{enumerate}
    \item Affected hubs ($SR$): For any $v$ from this set, a label $(v,\cdot,\cdot)$ should be renewed, deleted, or newly inserted. 
    \item Affected ordinary vertices ($R$): For any $v$ from this set, $L(v)$ may require update but $v$ is not an affected hub.
\end{enumerate}
}
\rev{For all affected ordinary vertices $v$, there is no need to renew, delete, or insert any label of the form $(v,\cdot,\cdot)$. Therefore, we can skip the BFS rooted at these vertices when updating the index. In addition, if $v$ is an affected hub, $L(v)$ may also require updates. Thus, we define the set of affected hubs as $\underline{S}ender$ $and$ $\underline{R}eceiver$ ($SR$) and the set of affected ordinary vertices as $\underline{R}eceiver$ $Only$ ($R$). During the decremental update, we only need to conduct BFS from vertices in the $SR$ set to update labels. It is worth noting that SPC-Index only counts those shortest paths in which one of the endpoints has the highest rank. Thus, we need to find out all such shortest paths that disappear or emerge due to the deletion of $(a,b)$. And the vertices in $SR$ are the highest-ranked endpoints of these shortest paths.}

\begin{definition} [$SR$ set]
    \rev{A vertex $v$ is considered in $SR$ if in one of the following two cases: For any arbitrary $u$, A) There exists at least one shortest path from $v$ to $u$ passing through $(a,b)$ where $v$ has the highest rank. Or B) All shortest paths from $v$ to $u$ pass through $(a,b)$. }
\label{def:sr}
\end{definition}

\rev{We can easily deduce that Condition A) is equivalence to the fact that $v$ is a hub of both $a$ and $b$, while Condition B) implies that $\texttt{spc}_i(v,a)=\texttt{spc}_i(v,b)$. In addition, For each $v\in SR$, we have either $\texttt{sd}_i(v,a)+1=\texttt{sd}_i(v,b)$ or $\texttt{sd}_i(v,b)+1=\texttt{sd}_i(v,a)$. Thus, $SR$ can be partitioned into $SR_a$ and $SR_b$ that contain the vertices closer to $a$ and $b$, respectively. }

\begin{figure}[thb]
\label{sr_example_1}
\centering
\includegraphics[scale=0.65]{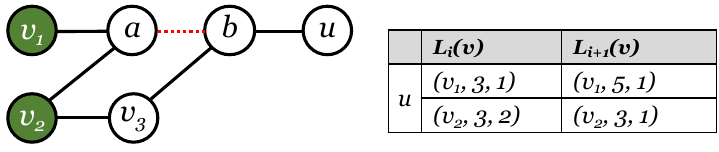} 
\caption{\rev{Examples of $SR$}}
\label{fig:sr_example}
\end{figure}

\begin{example}
    \rev{In Figure~\ref{fig:sr_example}, assume $v_1\leq v_2\leq v_3\leq a\leq b\leq u$. The green vertices represent two examples of $SR_a$ that satisfy Condition A). The table displays the changes in $u$'s label, with only the labels with hubs $v_1$ and $v_2$ being shown. After deleting $(a,b)$, the shortest path from $v_1$ to $u$ with $v_1$ as the highest-ranked node has changed from $v_1\rightarrow a\rightarrow b\rightarrow u$ to $v_1\rightarrow a\rightarrow v_2\rightarrow v_3\rightarrow b\rightarrow u$, resulting in a change in $u$'s label. For $v_2$, by losing one of the shortest paths to $u$, the label $(v_2,3,2)$ has changed to $(v_2,3,1)$. Both $v_1$ and $v_2$ are hubs of $a$ and $b$ and are part of $SR_a$. In Figure~\ref{fig:insert_example}, vertex $w$ is an example of $SR_a$ that satisfies Condition B). Even though $w$ is not a hub of $a$ or $b$. After deleting $(a,b)$, a shortest path from $w$ to $u$ with $w$ as the highest rank emerges which should be counted to the index.}
\end{example}

\begin{definition} [$R$ set]
    \rev{A vertex $v$ is in $R$ if for any arbitrary $u$. A) There exists at least one shortest path from $v$ to $u$ passing through $(a,b)$. And B) $v\notin SR$.}
\end{definition}

\rev{In other words, every vertex $v$ in $R$ satisfies $|\texttt{sd}_i(v,a)-\texttt{sd}_i(v,b)|=1$ but $v$ does not belong to $SR$. Consequently, we can also split $R$ into $R_a$ and $R_b$ in the same way. Note that the four sets $SR_a$, $SR_b$, $R_a$ and $R_b$ are disjoint, i.e., $SR_a\cap SR_b\cap R_a\cap R_b=\phi$.}

\begin{figure*}[htb]
	\begin{center}
        \subfigure[Delete Edge $(v_1, v_2)$.]{
			\label{dec_graph}
			\centering
			\includegraphics[scale=0.7]{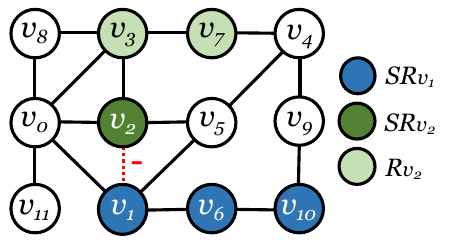} 
		}\hspace{2mm}
		\subfigure[Update Rooted at Hub $v_1$.]{
			\label{dec_1}
			\centering
			\includegraphics[scale=0.7]{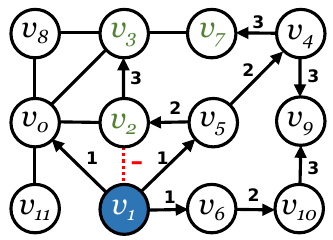} 
		}\hspace{2mm}
		\subfigure[Update Rooted at Hub $v_2$.]{
			\label{dec_2}
			\centering
			\includegraphics[scale=0.7]{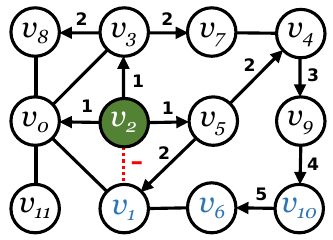}     
		}\vspace{-1mm}
        \subfigure[Update Steps for Affected Hubs $v_1$ and $v_2$]{ \footnotesize
            \label{dec_1_table}
            \centering
            \begin{tabular}{c||c|c|c|c||c|c|c}
              \hline
              \makecell[l] {\emph{BFS Iteration}} & \makecell[l] {\emph{Meeting Vertex}} &
              \makecell[l] {\emph{Rank Pruning}} & \makecell[l] {\emph{Distance Pruning}} &
              \makecell[l] {$SR_{v_2}\cup R_{v_2}$} & \makecell[l] {\emph{Existing Label}} &
              \makecell[l] {\emph{D[v], C[v]}} & \makecell[l] {\emph{Update Operation and New Label}}\\\hline
              \multicolumn{8}{c}{Affected Hub $v_1$}\\\hline
              \multirow{2}*{D=1} & $v_0$  & Pruned &     -        &  -  &      -      &  -  & -                  \\\cline{2-8}
              ~                  & $v_5$ $v_6$ & Pass   & Pass ($\Bar{d}$=1=D)   & No  &      -      &  -  & -                  \\\hline
              \multirow{2}*{D=2} & $v_2$  & Pass   & Pass (2=D)   & Yes & ($v_1,1,1$) & 2,1 & Renew to ($v_1,2,1$)  \\\cline{2-8}
              ~                  & $v_4$ $v_{10}$ & Pass   & Pass (2=D)   & No  &      -      &  -  & -                  \\\hline
              \multirow{3}*{D=3} & $v_3$  & Pass   & Pruned (2<D) & Yes  & ($v_1,2,1$) &  -  & Delete ($v_1,2,1$) in label removal process\\\cline{2-8}
              ~                  & $v_7$  & Pass   & Pass (3=D)   & Yes & ($v_1,3,2$) & 3,1 & Renew to ($v_1,3,1$)  \\\cline{2-8}
              ~                  & $v_9$  & Pass   & Pass ($\infty$>D)   & No  &      -      &  -  & -                  \\\hline
              \multicolumn{8}{c}{Affected Hub $v_2$, steps with D$\leq$3 are omitted}\\\hline
              D=4                &$v_{10}$& Pass   & Pass (4=D)   & Yes &  No Label   & 4,1 & Insert ($v_2,4,1$) \\\hline
              D=5                & $v_6$  & Pass   & Pruned (3<D) &  -  &      -      &  -  & -                  \\\hline
            \end{tabular}
        }
	\end{center}
    \vspace{-4mm}
	\caption{Example of Decremental Update. Delete Edge $(v_{1}, v_{2})$ from $G$.} \label{fig:dec_example}
    \vspace{-2mm}
\end{figure*}

\begin{example}
    A full example of $SR$ and $R$ is shown in \autoref{fig:dec_example}(a). Assume edge $(v_1, v_2)$ is deleted. We have $SR_{v_1}=\{v_1, v_6, v_{10}\}$ where $v_1\in L_i(v_1)\cap L_i(v_2)$, while $v_6$ and $v_{10}$ satisfy $\texttt{spc}(\cdot,v_1)=\texttt{spc}(\cdot,v_2)$. $SR_{v_2}=\{v_2\}$, and $R_{v_2}=\{v_3, v_7\}$. There is no element in $R_{v_1}$.
    \label{srr}
\end{example}

\begin{lemma}
    Given a graph $G_i$ with the SPC-Index $L_i$. Assume edge $(a,b)$ is deleted from $G_i$ and the resulting graph is $G_{i+1}$. For any $v,u\in V$, $(v,\cdot,\cdot)\in L_i(u)$ may require update or $(v,\cdot,\cdot)$ may be newly inserted into $L_i(u)$ only if $v\in SR_a$ and $u\in SR_b\cup R_b$ or $v\in SR_b$ and $u\in SR_a\cup R_a$.
    \label{lemma:label_update}
\end{lemma}

\rev{The above lemma provides us with a way to identify updates with the assistance of $SR$ and $R$. To avoid reconstructing the entire index, our approach involves conducting a BFS from the affected hubs in $SR$, detecting and updating outdated labels, or complementing new labels simultaneously. Our decremental algorithm works in two main phases as follows:
\renewcommand{\labelenumii}{\arabic{enumi}.\arabic{enumii}}
\begin{enumerate}
    \item Compute $SR_a$, $SR_b$, $R_a$ and $R_b$
    \item Search and update labels
\end{enumerate}}

In what follows, we describe the main steps of our decremental update algorithm, \textsc{DecSPC}, followed by presenting the two main phases in detail.

\begin{algorithm}[htb]
    $SR_a, SR_b, R_a, R_b \leftarrow $ \textsc{Srr}SEARCH($G_i, a, b$)\;
    $SR \leftarrow $ \textbf{sort}($SR_a \cup SR_b$)\;
    $G_{i+1} \leftarrow G_{i} \ominus (a,b)$\;
    $L_{ab} \leftarrow \{h | h\in L_i(a)\cap L_i(b)\}$\;
    \For{{\bf each} $h \in SR$ {\rm in descending order}} {
        \If{$h \in SR_a$} {
            $H_a \leftarrow True$ {\bf if} $h\in L_{ab}$\;
            \textsc{Dec}UPDATE($h, SR_b, R_b, H_a$)\;
        } \Else {
            $H_b \leftarrow True$ {\bf if} $h\in L_{ab}$\;
            \textsc{Dec}UPDATE($h, SR_a, R_a, H_b$)\;
        }
    }
\caption{\textsc{Dec}SPC($G_i,L_i,a,b$)}
\label{alg:dec_spc}
\end{algorithm}

\rev{Algorithm \ref{alg:dec_spc} outlines the main steps of \textsc{DecSPC}. Firstly, in line 1, $SR_a$, $SR_b$, $R_a$, and $R_b$ are computed using \textsc{SrrSEARCH}. Next, $SR_a$ and $SR_b$ are merged into $SR$ and sorted by vertices' ranks (line 2). The graph $G_{i+1}$ is then obtained by deleting edge $(a,b)$ from $G_i$ (line 3), and the common hubs between $a$ and $b$ are stored in $L_{ab}$ in line 4. Starting from each affected hub $h$ in descending order of rank, the decremental updates are conducted in lines 5-11. Depending on the $SR$ set that $h$ belongs to, the update algorithm \textsc{DecUPDATE} is initialized with corresponding parameters. A boolean flag, $H_a$ (line 7) or $H_b$ (line 10), is used to determine whether $h$ is in $L_{ab}$ and trigger the label removal process if true in \textsc{DecUPDATE}.}

\subsubsection{\rev{Compute $SR_a$, $SR_b$, $R_a$ and $R_b$}}~\label{srr_a_to_b}
\rev{Algorithm \ref{alg:srr_search} outlines the procedure of computing $SR_a$ and $R_a$ in a single search. To compute $SR_b$ and $R_b$, we can modify the algorithm by replacing $SR_a$, $R_a$, and \textsc{SpcQuery($v,a$)} with $SR_b$, $R_b$, and \textsc{SpcQuery($v,b$)}, respectively.}

\begin{algorithm}[htb]
    $L_{ab} \leftarrow \{h | h\in L_i(a)\cap L_i(b)$\}\;
    $SR_a \leftarrow \emptyset$; $SR_b \leftarrow \emptyset$;
    $R_a \leftarrow \emptyset$; $R_b \leftarrow \emptyset$\;
    \For{{\bf each} $u \in V$} {
        $D[u] \leftarrow \infty$; $C[u] \leftarrow 0$\;
    }
    $D[a] \leftarrow 0$; $C[a] \leftarrow 1$\;
    Queue $Q \leftarrow $ $\emptyset$; $Q.{\rm enqueue}(a)$\;
    \While{$Q$ {\rm $\neq \emptyset$}}{
        $v \leftarrow Q.{\rm dequeue}()$\;
        $d, c \leftarrow $\textsc{Spc}Query$(v,b)$\;
        {\bf if} $D[v] + 1 \not= d$ {\bf then continue}\;
        \If{$v \in L_{ab}$ {\bf or} $C[v] = c$} {
            $SR_a \leftarrow SR_a \cup \{v\}$\;
        } \Else {
            $R_a \leftarrow R_a \cup \{v\}$\;
        }
        \For{{\bf each} $w \in \texttt{nbr}(v)$}{
            \If{$D[w] = \infty$}{
                $D[w] \leftarrow D[v] + 1$; $C[w] \leftarrow C[v]$\;
                $Q.{\rm enqueue}(w)$\;
            } \ElseIf{$D[w] = D[v] + 1$}{
                $C[w] \leftarrow C[w] + C[v]$\;
            }
        }
    }
\caption{\textsc{Srr}SEARCH($G_i, a, b$)}
\label{alg:srr_search}
\end{algorithm}

\rev{Algorithm \ref{alg:srr_search} implements a BFS starting from $a$ and prunes when meeting an unaffected vertex. We use a queue $Q$ to keep track of the visited vertices, and the shortest distance and the shortest path counting from $a$ for all visited vertices are stored in arrays $D[\cdot]$ and $C[\cdot]$, respectively. Line 1 identifies the common hubs of $a$ and $b$ and saves them into $L_{ab}$. Lines 2 through 6 initialize $SR_a$, $SR_b$, $R_a$, and $R_b$, $D[\cdot]$ and $C[\cdot]$, and $Q$. The BFS is executed in lines 7 through 20. At each visited vertex $v$, the current index $L_i$ is used to compute $\texttt{sd}_i(v,b)$ and $\texttt{spc}_i(v,b)$, which are stored in $d$ and $c$, respectively (line 9). Unaffected vertices are pruned at line 10, where a vertex $v$ is pruned if $\texttt{sd}_i(v,a)+1\not= \texttt{sd}_i(v,b)$. Then, vertex $v$ is added to $SR_a$ based on the conditions specified in its definition as follows.}

\begin{enumerate} [leftmargin=*]
    \item Condition A implies that $\texttt{spc}_i(\Hat{v},a)>0$ and $\texttt{spc}_i(\Hat{v},b)>0$, i.e., $v$ should be the hub of both $a$ and $b$. Thus, we check $v\in L_{ab}$.
    \item Condition B indicates that $\texttt{spc}_i(v,a)=\texttt{spc}_i(v,b)$.
\end{enumerate}

\rev{Vertices that satisfy the above conditions are added to $SR_a$ (lines 11-12) while others are added to $R_a$ (lines 13-14). At the end of the BFS, $SR_a$ and $R_a$ are completed. The algorithm can then be repeated with the aforementioned modification to find $SR_b$ and $R_b$.}

\subsubsection{\rev{Search and Update Labels}}
\rev{In this section, we present the second phase of \textsc{DecSPC}, and the corresponding algorithm is shown in Algorithm \ref{alg:dec_update}. Firstly, we introduce a modified query algorithm called \textsc{PreQUERY}($s,t$). The only difference between \textsc{PreQUERY}($s,t$) and \textsc{SpcQUERY}($s,t$) is the addition of the line {\bf if} $h=s$ {\bf then break}\; after line 2. \textsc{PreQUERY}($s,t$) returns the shortest distance $\Bar{d}$ and shortest path counting $\Bar{c}$ between $s$ and $t$ by utilizing the hubs with ranks higher than $s$, where $\Bar{d}$ serves as an upper bound on the shortest distance between $s$ and $t$.}

\begin{algorithm}[htb]
    \For{{\bf each} $u \in V$} {
        $D[u] \leftarrow \infty$; $C[u] \leftarrow 0$; $U[u] \leftarrow$ {\bf False}\; 
    }
    $D[h] \leftarrow 0$; $C[h] \leftarrow 1$\;
    Queue $Q \leftarrow $ $\emptyset$; $Q.{\rm enqueue}(h)$\;
    \While{$Q$ {\rm $\neq \emptyset$}}{
        $v \leftarrow Q.{\rm dequeue}()$\;
        $\Bar{d}, \Bar{c} \leftarrow $\textsc{Pre}QUERY$(h,v)$\;
        {\bf if} $\Bar{d} < D[v]$ {\bf then continue}\;
        \If{$v\in SR\cup R$} {
            \If{$h\notin L(v)$} {
                Insert $(h,D[v],C[v])$ to $L(v)$\;
            } \Else {
                $(h,d,c) \leftarrow$ label with hub $h$ from $L(v)$\;
                \If{$d\not=D[v]$ {\bf or} $c\not=C[v]$} {
                    Replace $(h,d,c)$ with $(h,D[v],C[v])$\;
                }
            }
            $U[v] \leftarrow $ {\bf True}\;
        }
        
        \For{{\bf each} $w \in \texttt{nbr}(v)$}{
            \If{$D[w] = \infty$ {\bf and} $h\leq w$}{
                $D[w] \leftarrow D[v] + 1$; $C[w] \leftarrow C[v]$\;
                $Q.{\rm enqueue}(w)$\;
            } \ElseIf{$D[w] = D[v] + 1$}{
                $C[w] \leftarrow C[w] + C[v]$\;
            }
        }
    }
    \If{$H_{ab} =$ {\bf True}} {
        \For{{\bf each} $u\in SR\cup R$} {
            \If{$U[u] =${\bf False and} $h\in L(u)$} {
                Remove $(h,d,c)$ from $L(u)$\;
            }
        }
    }
\caption{\textsc{Dec}UPDATE($h, SR, R, H_{ab}$)}
\label{alg:dec_update}
\end{algorithm}

\rev{To update the index, we begin by processing each affected hub in descending order of rank. Algorithm \ref{alg:dec_update} takes four input parameters: the current affected hub $h$, the opposite $SR$ and $R$ sets, and a flag indicating whether $h$ is the common hub of $a$ and $b$. The algorithm utilizes a queue $Q$ to track visited vertices, and arrays $D[\cdot]$ and $C[\cdot]$ to store BFS distance and path counting. Additionally, a Boolean array $U[\cdot]$ is introduced to record the visited and updated status of each vertex, with $U[v]$ being True if $v$ has been visited and updated, and False otherwise.
Lines 1 to 4 handle the initialization process. The BFS begins from $h$ to identify the vertices whose $(h,\cdot,\cdot)$ labels require updating (lines 5-22). During the BFS, vertex $v$ is pop from $Q$, and \textsc{PreQUERY}($h,v$) returns $\Bar{d}$ and $\Bar{c}$ (lines 6-7). The BFS is pruned if $\Bar{d}<D[v]$, indicating that the existing index already covers all the shortest paths between $h$ and $v$ (line 8). Otherwise, if $v \in SR \cup R$, a new label is inserted into $L(v)$ if $h$ is not the hub of $v$ (lines 10-11). Alternatively, if there are changes in the distance or shortest path counting, the label is updated accordingly (lines 12-15). At this point, the label $(h,\cdot,\cdot) \in L(v)$ is up-to-date, and $U[v]$ is set to True (line 16). The BFS continues with ranking pruning constraints (lines 17-22).
After completing the updates and insertions, if $h$ is the common hub of $a$ and $b$, labels $(h,\cdot,\cdot) \in L(u)$ where $u \in SR \cup R$ and $U[u]$ is False are considered for removal (lines 23-26). This is because either $h$ and $u$ have become disconnected or the label is dominated by other labels.}

\begin{example}
    \rev{Continuing from Example~\ref{srr}, we merge $SR_{v_1}$ and $SR_{v_2}$ to obtain $SR=\{v_1, v_2, v_6, v_{10}\}$. Figure~\ref{fig:dec_example}(b) and (c) illustrate the steps of the decremental update rooted at $v_1$ and $v_2$, respectively. The numbers next to the edges indicate the shortest distance from the current affected hub. The table in Figure~\ref{fig:dec_example}(d) provides detailed update information.
    We begin by performing a BFS from $v_1$ as depicted in Figure~\ref{fig:dec_example}(b). During the first round of BFS (i.e., $D=1$), $v_0$ is pruned directly since $v_0\leq v_1$. For $v_5$ and $v_6$, although their $\Bar{d}$ values are both 1, which equals $D$, they are not in $SR_{v_2}\cup R_{v_2}$, so their labels are unaffected, and the BFS continues.  When $D=2$ and we encounter $v_2$ from $SR_{v_2}\cup R_{v_2}$, $\Bar{d}=2=D$, along with the existing label $(v_1,1,1)$, indicating the emergence of a new shortest path from $v_1$ to $v_2$ with $v_1$ as the highest rank, i.e., $v_1\rightarrow v_5\rightarrow v_2$. Thus, the existing label needs to be updated to $(v_1,2,1)$. When $D=3$ and we reach $v_3$, $\Bar{d}=2<D$, indicating that there are no shortest paths from $v_0$ to $v_3$ with $v_0$ as the highest rank. Consequently, the existing label in $L(v_3)$ should be deleted. However, we postpone the deletion at this stage, and instead, keep $U[v_3]$ as False until the BFS concludes. This enables the label to be removed during the label cleaning step (lines 23-26 in Algorithm~\ref{alg:dec_update}). Note that the BFS is pruned at $v_3$, which means we don't need to visit the subsequent vertices (if any) whose labels should be deleted, as their $U[\cdot]$ values remain False. When encountering $v_7$, its label changes from $(v_1,3,2)$ to $(v_1,3,1)$. For the affected hub $v_2$, the only update to the index is the insertion of a new label $(v_2,4,1)$ for $v_{10}$, resulting from the new shortest path $v_2\rightarrow v_5\rightarrow v_4\rightarrow v_9\rightarrow v_{10}$. The remaining steps for $v_6$ and $v_{10}$ are omitted as there is no change to the index by these hubs.}
\end{example}

\rev{The example above illustrates the effectiveness of \textsc{DecSPC} in achieving the following objectives: 1) Limiting the number of BFS operations for updates by carefully selecting affected hubs. 2) Accurately identifying the vertices whose labels require updating. In the given example, all such vertices are found within $SR_{v_2} \cup R_{v_2}$ for the affected hub $v_1$. 3) Avoiding the removal of all potentially affected labels initially, as done by existing methods for SD-Index. Instead, our algorithm efficiently removes redundant labels after the main update step, thereby improving efficiency.}

\subsubsection{Isolated Vertex Deletion}
Given a graph $G_i$, edge $(a,b)$ is deleted from $G_i$ and the resulting graph is $G_{i+1}$. Assume $b$ is the vertex whose degree is one in $G_i$, i.e., $\texttt{deg}_i(b)=1$. If both $\texttt{deg}_i(a)=\texttt{deg}_i(a)=1$, let $b$ be the one with lower rank. According to the vertex ordering and the above assumption, we have $a\leq b$. In this case, after deleting $(a,b)$, $b$ is disconnected from other vertices in the graph, thus we can empty $L(b)$ and just leave $(b,0,1)$. Because $a\leq b$, $\texttt{spc}_i(\Hat{b},v)=0$ for all $v\in G_i\backslash \{b\}$. No label with $b$ as hub exists in other vertices' label set so that no more operations are required. This optimization can avoid BFSs from all affected hubs.

\begin{theorem} [Correctness]
    Given $G_{i+1} = G_i \ominus (a,b)$, $L_i$ is the SPC-Index of $G_i$ and $L_{i+1}$ is the updated SPC-Index by Algorithm \ref{alg:dec_spc}. $L_{i+1}$ obeys ESPC regarding $G_{i+1}$.
    \label{theorem:dec_correct}
\end{theorem}

\begin{theorem} [Time Complexity]
    Assume average label size $|L(v)|=O(l)$ for any vertex $v$, the sizes of $SR_a$ and $SR_b$ are $O(s_a)$, $O(s_b)$, respectively. The sizes of $SR_a\cup R_a$ and $SR_b\cup R_b$ are $O(S_a)$, $O(S_b)$, respectively. BFS visits $O(k)$ vertices on average for each affected hub. The time complexity of Algorithm \ref{alg:dec_spc} is $O((S_a+S_b)l+(k+S_b)ls_a+(k+S_a)ls_b)$.
    \label{theorem:dec_time}
\end{theorem}